\documentclass[12pt]{article} 
 \pdfoutput=1

\usepackage{amsmath,amssymb,amsfonts}
\usepackage{psfrag}
\usepackage{color}
\definecolor{darkblue}{rgb}{0.1,0.1,.7}
\usepackage[colorlinks, linkcolor=darkblue, citecolor=darkblue, urlcolor=darkblue, linktocpage]{hyperref} 
\usepackage[square, comma, sort&compress,numbers]{natbib}
\usepackage[]{amsmath}
\usepackage[]{graphicx}
\usepackage[]{latexsym}
\usepackage{geometry}
\usepackage{amscd}
\usepackage{braket}
\usepackage[all,cmtip]{xy}
\usepackage{mathrsfs}
\usepackage[margin=10pt,font=small,labelfont=bf]{caption}
\usepackage{simplewick}
\usepackage{bbold}
\usepackage{changepage}
\usepackage{dsfont}
\usepackage{bm}
\usepackage{float}
\usepackage[utf8]{inputenc}
\usepackage{setspace}

\numberwithin{equation}{section}

\newcommand{\be}{\begin{eqnarray}}
\newcommand{\ee}{\end{eqnarray}}
\newcommand{\bea}{\begin{eqnarray}}
\newcommand{\eea}{\end{eqnarray}}

\renewcommand \Re    {\mathop{\mathrm{Re}}}
\renewcommand \Im    {\mathop{\mathrm{Im}}}

\newcommand   \De {\Delta}

\newcommand   \f  {\phi}
\newcommand   \p  {\psi}

\newcommand   \s  {\sigma}

\def\beq{\begin{equation}} 
\def\eeq{\end{equation}} 
\def\<{\langle}
\def\>{\rangle}

\def\nn{\nonumber} 
 
\def\cO {{\cal O}}

\begin{document}

\vspace*{-.6in} \thispagestyle{empty}
\begin{flushright}
\end{flushright}
\vspace{.2in} {\Large
\begin{center}
{\bf Conformal Bootstrap in the Regge Limit \vspace{.1in}}
\end{center}
}
\vspace{.2in}
\begin{center}
{\bf 
Daliang Li$^{a}$, David Meltzer$^b$,  David Poland$^{b}$}
\\
\vspace{.2in} 
$^a$ {\it  Department of Physics and Astronomy, Johns Hopkins University, Baltimore, MD 21218}\\
$^b$ {\it Department of Physics, Yale University, New Haven, CT 06511}\\
\end{center}

\vspace{.2in}

\begin{abstract}

We analytically solve the conformal bootstrap equations in the Regge limit for large $N$ conformal field theories. For theories with a parametrically large gap, the amplitude is dominated by spin-2 exchanges and we show how the crossing equations naturally lead to the construction of AdS exchange Witten diagrams. We also show how this is encoded in the anomalous dimensions of double-trace operators of large spin and large twist. We use the chaos bound to prove that the anomalous dimensions are negative. Extending these results to correlators containing two scalars and two conserved currents, we show how to reproduce the CEMZ constraint that the three-point function between two currents and one stress tensor only contains the structure given by Einstein-Maxwell theory in AdS, up to small corrections. Finally, we consider the case where operators of unbounded spin contribute to the Regge amplitude, whose net effect is captured by summing the leading Regge trajectory. We compute the resulting anomalous dimensions and corrections to OPE coefficients in the crossed channel and use the chaos bound to show that both are negative. 

\end{abstract}

\newpage

\setcounter{tocdepth}{2}
\begin{spacing}{.2}
\tableofcontents
\end{spacing}

\newpage

\section{Introduction}

The bootstrap program, which seeks to study quantum field theories by studying their general consistency conditions, has shown to be remarkably powerful when applied to conformal field theories (CFTs)~\cite{Ferrara:1973yt,Polyakov:1974gs}. By analyzing the constraints of unitarity and crossing symmetry, the modern bootstrap program has yielded new insights into the structure of CFTs \cite{Rattazzi:2008pe}. With the bootstrap, it is possible to rigorously study the space of CFTs with a small number of relevant operators \cite{ElShowk:2012ht,El-Showk:2014dwa,Kos:2014bka,Simmons-Duffin:2015qma, Kos:2016ysd}. 

Another prominent application of the bootstrap program is in the study of quantum gravity. Through the holographic principle, one can unambiguously define a theory of quantum gravity living in an asymptotically AdS$_{d+1}$ spacetime in terms of a CFT living on its boundary. Various observables in the gravity theory can be translated into observables of the boundary CFT, which can then be studied using symmetry and consistency conditions of the CFT. The power of this approach is in its generality. One is not necessarily constrained to a particular incarnation of quantum gravity or the usual limit of large $N$ or large gap. Instead, very general results can be obtained based on symmetries and consistency conditions that apply to all gravity theories dual to boundary CFTs. For example, the gravitational interaction between objects in AdS at superhorizon distances has been shown to be attractive for a wide range of quantum gravitational theories ~\cite{Fitzpatrick:2012yx,Komargodski:2012ek,Li:2015rfa,Li:2015itl,Hofman:2016awc}. In AdS$_3$, this approach has also unveiled non-perturbative effects that are crucial in solving the information loss problem of black holes that are difficult to resolve using traditional methods~\cite{Fitzpatrick:2016ive, Dyer:2016pou,Chen:2017yze}.

In this work, we will use analytic bootstrap techniques to understand general constraints on large $N$ CFTs and their AdS duals. It is conjectured that any large $N$ CFT with a parametrically large gap in its higher spin single-trace sector is well described by a local bulk dual of Einstein gravity plus matter. This is an extremely non-trivial statement from the CFT point of view, since it suggests that by taking $N$ and $\Delta_{gap}$ large, an infinite amount of CFT data is uniquely fixed to be the value corresponding to Einstein gravity in the bulk. The bootstrap approach is particularly suitable for investigating this phenomenon and has already provided strong evidence for the conjecture. In the study of external scalar operators, it has been shown that there is a one-to-one map between homogeneous solutions to crossing and local, quartic interactions in AdS~\cite{Heemskerk:2009pn,Heemskerk:2010ty}.  Similar counting also works for exchange Witten diagrams. For operators with spin, it has been shown that the three-point function of stress tensors $\<TTT\>$ \cite{Afkhami-Jeddi:2016ntf} is fixed to take the form predicted by Einstein gravity.\footnote{The result in \cite{Afkhami-Jeddi:2016ntf} is based on certain assumptions about the contributions of double-trace states in the Regge limit. We will clarify the role of such contributions in the present work.}  It has also been shown that higher derivative interactions are suppressed by $\Delta_{gap}$, consistent with expectations from effective field theory~\cite{Caron-Huot:2017vep} (see also~\cite{Alday:2014tsa,Alday:2016htq,Aharony:2016dwx,Rastelli:2016nze,Perlmutter:2016pkf}).

We will study the universality of Einstein gravity, as well as regimes beyond it, by solving the bootstrap equations in the Regge limit. In flat space scattering, the Regge limit corresponds to taking $|s|\gg |t|$, with $t$ held fixed. This limit is sensitive to the spectrum of higher spin particles. Among the various states exchanged by the two scattering particles, the dominant contribution to the amplitude comes from the tower of particles with the lowest mass for each spin, or the leading Regge trajectory. In CFTs, there is an analogous kinematic limit of four-point functions that is dual to two-to-two, high energy, fixed impact parameter scattering in the AdS bulk \cite{Brower:2006ea,Cornalba:2007fs,Cornalba:2007zb,Cornalba:2006xk,Cornalba:2006xm}. In CFTs the leading Regge trajectory corresponds to the tower of single-trace operators exchanged in the $\psi\psi$ channel with the lowest scaling dimension at each spin \cite{Costa:2012cb}. The Regge limit has also seen renewed attention given its connection to chaos, and we will use the recently studied bound on chaos \cite{Maldacena:2015waa} to constrain the space of large $N$ CFTs in $d>2$.

Our key observables are the anomalous dimensions and OPE coefficients of the double-trace operators $[\f\psi]_{n,j}\sim \f\partial^{\mu_{1}}...\partial^{\mu_{j}}\partial^{2n}\psi$ with spin $\ell$ and dimension $\Delta_\phi+\Delta_\psi+2n+j$. The anomalous dimensions are the first large $N$ corrections to these canonical dimensions. In the Regge limit, they correspond to the bulk phase shift of the scattering particles,
\be
\gamma_{n,j}=-\frac{1}{\pi} \delta(s,b),
\label{eq:phaseshiftIntro}
\ee
where the bulk center of mass energy $\sqrt{s}$ and impact parameter $b$ are controlled by $n$ and $\ell$ as in (\ref{eq:phaseshift}). We compute $\gamma_{n,j}$ by solving the bootstrap equations. 

In the first part of this work, we will take $s\ll\Delta_{gap}^{2}$. In this regime we indeed recover Einstein gravity. In particular, the anomalous dimensions we compute agree with the phase shifts obtained from local bulk gravity~\cite{Camanho:2014apa}. For scalar 4-point functions, we argue that these anomalous dimensions are negative using the chaos bound. We also sketch the generalization of this argument to correlation functions containing currents. The resulting condition implies that the $\langle JJT\rangle$ three point function only contains the structure obtained from Einstein-Maxwell theory in the bulk, up to small corrections.

In the second part of this work, we ramp up the scattering energy to probe beyond the scale set by $\Delta_{gap}$. This is interesting both from the bulk and boundary point of view. Large $N$ CFTs with a parametrically large gap in the single-trace spectrum are expected to be non-generic. This truncation occurs in planar, $\mathcal{N}=4$ SYM at strong coupling, but the consequences of crossing symmetry remain elusive when we are away from this limit and infinite towers of higher spin single-trace operators, organized into Regge trajectories, contribute to the correlation function. In the dual AdS theory, we would like to understand whether universality exists beyond the gravity limit. In particular, is string theory the only possible UV completion of Einstein gravity? From the study of flat space scattering amplitudes in weakly coupled theories, there is convincing evidence that the presence of higher spin particles implies some general stringy properties \cite{Caron-Huot:2016icg,Cardona:2016ymb}. In this paper, we will show that with mild assumptions on the Regge trajectory, string-like behavior does emerge in the CFT. Understanding the properties of the leading Regge trajectory is also crucial in resolving causality problems for weakly coupled theories of gravity \cite{Camanho:2014apa}.

In addition to computing the anomalous dimensions and OPE coefficients of the double-trace operators $[\f\psi]_{n,j}$, we also constrain their signs using unitarity of the boundary CFT, both in the gravity regime and when there is an exchange of an entire Regge trajectory. Previously, similar sign constraints were discovered in the study of the lightcone bootstrap~\cite{Fitzpatrick:2012yx,Komargodski:2012ek} for external operators with spin~\cite{Li:2015itl,Hofman:2016awc}, where the exchange of the stress tensor only leads to negative anomalous dimensions if the conformal collider bounds~\cite{Hofman:2008ar} are satisfied. This negativity implies the attractiveness of bulk gravity at long distances. Using CFT axioms, it is possible to prove the collider bounds~\cite{Hartman:2015lfa, Hartman:2016dxc, Komargodski:2016gci, Hofman:2016awc} and the averaged null energy condition (ANEC) more broadly \cite{Hartman:2016lgu,Faulkner:2016mzt}. We will see similar behavior in the Regge limit. We prove the negativity of the anomalous dimensions in a wide range of theories. This result is explicitly connected to the causality of the bulk gravitational theory through (\ref{eq:phaseshiftIntro}). We also prove that corrections to the OPE coefficients of $[\f\psi]_{n,j}$ are negative. This has a simple bulk interpretation in terms of AdS unitarity. 

\begin{it}
Note added: after this work was completed, \cite{Alday:2017gde} and \cite{KPZ2017} appeared, which also consider the Regge limit in CFTs and have some overlap with our work.
 \end{it}

\subsection{Summary of Results}

Our paper is organized as follows. In section \ref{sec:largeN} we will introduce some useful notation and conventions for the study of large $N$ CFTs. For simplicity we will present known results for the four-point function of scalar operators $\<\f\p\p\f\>$ in the $\f\f\rightarrow \p\p$ Regge limit. With a precise definition of the Regge limit as an analytic continuation of the Euclidean four-point function, we will show that this limit naturally isolates operators with large spin.

In section \ref{sec:GravityLimit} we will study the bootstrap equations in the Regge limit for CFTs where the single-trace spectrum is bounded in spin. The main application will be for CFTs dual to an AdS theory of gravity plus matter, but we will start more generally and consider the exchange of an isolated, spin-$j$, single-trace operator $\mathcal{O}_{\Delta,j}$. We show how, with some simple assumptions, solving the crossing equations when a single-trace operator is exchanged in the $\f\f\rightarrow \mathcal{O}\rightarrow \p\p$ channel naturally leads to the construction of AdS exchange Witten diagrams. We then move on to consider the four-point function $\<J\f\f J\>$, where $J$ is a conserved current and $\f$ is a scalar. We show the double-trace anomalous dimensions agree exactly with the phase shifts calculated in AdS for the scattering of a gauge boson through a shock wave \cite{Camanho:2014apa}. Finally, we will use the chaos bound to show that the large $n$ and $j$ anomalous dimensions must be negative, or that the dual AdS theory is causal. 

The study of these classes of double-trace operators for external scalars has been considered in the past for large $N$ theories, using the lightcone bootstrap in \cite{Kaviraj:2015xsa,Kaviraj:2015cxa}, for the exchange of scalar single-trace operators in $d=2$ using twist blocks in \cite{Alday:2016njk}, and using impact parameter partial waves inspired by AdS/CFT in \cite{Cornalba:2006xm,Cornalba:2006xk,Cornalba:2007zb}. For these theories we will recover the results of \cite{Cornalba:2006xm,Cornalba:2006xk,Cornalba:2007zb} and provide new evidence that their impact parameter partial waves correspond to conformal blocks in the appropriate limit. We give more details on the connection between conformal blocks and the impact parameter waves in appendix \ref{App:Impact}. By working directly in the Regge limit we will also simplify the derivation of the large $n$ and $j$ anomalous dimensions in comparison to the lightcone work.

In section \ref{sec:ReggeExchange} we will study the crossing symmetry equations when an entire Regge trajectory with an infinite number of operators is exchanged in the $\f\f\rightarrow \p\p$ channel. We will use the work of \cite{Costa:2012cb} to write down the form of the full correlation function and derive new results for the anomalous dimensions and corrections to the OPE coefficients. We propose an addition to the holographic dictionary, where the anomalous dimensions and correction to the OPE coefficients correspond to the real and imaginary part of the bulk phase shift, respectively. Requiring that crossing symmetry is satisfied on the first sheet will imply both new bounds on the t-channel OPE coefficients and the presence of an infinite number of new single-trace operators at tree level. Finally, using the chaos bound we will study constraints on the Regge intercept $j(0)$ and the phase of the correlation function in the Regge limit. This latter bound will imply that in theories like $\mathcal{N}=4$ SYM, where $1\leq j(0)\leq 2$, both anomalous dimensions and corrections to OPE coefficients must be negative. These bounds ensure the AdS dual is causal and obeys unitarity \cite{Cornalba:2008qf}.

\section{Regge Limit in Large N CFTs}
\label{sec:largeN}
We start by considering the CFT four-point function of two pairs of identical scalars
\bea
G(z,\bar{z})=\<\f(0)\p(z,\bar{z})\p(1)\f(\infty)\>,
\eea
where we have used conformal invariance to place the operators at the specified positions. The restriction to scalars is primarily to simplify the presentation (everything discussed in this section can be readily generalized to external spinning operators). Equating the s- and t-channel conformal block decompositions of this function yields the crossing symmetry equation:
\begin{small}
\begin{align}
\hspace{-.65cm}((1-z)(1-\bar{z}))^{-\Delta_{\p}}(z\bar{z})^{\frac{1}{2}(\Delta_{\f}+\Delta_{\p})}&\sum_{\mathcal{O}}\left(-\frac{1}{2}\right)^{j}C_{\f\f\mathcal{O}}C_{\psi\psi\mathcal{O}}g^{0,0}_{\Delta, j}(1-z,1-\bar{z}) \nonumber\\
= &\sum_{\mathcal{O'}}\left(-\frac{1}{2}\right)^{j'}C_{\f\psi\mathcal{O}'}C_{\psi\f\mathcal{O}'}g^{a,a}_{\Delta', j'}(z,\bar{z}),  \label{eq:Crossing}
\end{align}
\end{small}
where $a=\frac{1}{2}(\Delta_{\p}-\Delta_{\f})$. Our conventions for the conformal blocks are described in appendix \ref{App:Conventions}. In the s-channel, or $\f\p\rightarrow\p\f$ channel, it will be convenient to use the notation
\begin{align}
P_{\mathcal{O}} \equiv \left(-\frac{1}{2}\right)^{j}C_{\f\psi\mathcal{O}}C_{\psi\f\mathcal{O}} = \left(\frac{1}{2}\right)^{j}(C_{\f\psi\mathcal{O}})^2,
\end{align}
so $P_{\mathcal{O}}$ is a manifestly positive quantity.
It will also be useful to parametrize the cross ratios as: 
\bea
1-\bar{z}=\eta \s, \qquad 1-z=\s ,
\eea
with $\s>0$ and $0\leq\eta\leq1$. 

Then the t-channel Euclidean OPE limit is defined by taking $\s\rightarrow 0$ with $\eta$ fixed. In this limit each t-channel conformal block scales like 
\bea
g^{0,0}_{\Delta,j}(\s,\eta) \sim \s^{\Delta}\eta^{\frac{\Delta-j}{2}},
\eea
and the correlation function is dominated by operators of low dimension. We can also consider the lightcone limit by taking $\eta\rightarrow 0$ with $\s$ held fixed. In this limit operators of low twist $\tau=\Delta-j$ dominate. In both regimes we stay on the first sheet of $G(z,\bar{z})$ such that both its s- and t-channel conformal block decompositions converge. 

Unlike the above limits, the Regge limit is only defined on the second sheet. We will take $z$ around the origin, $z\rightarrow e^{-2\pi i}z$, and then take $z,\bar{z}\rightarrow 1$ at a fixed rate, or $\s\rightarrow 0$ for fixed $\eta$. On the second sheet the t-channel OPE is no longer convergent, but the s- and u-channel OPEs will remain convergent. Nevertheless, we can understand the physics of the Regge limit by studying individual t-channel conformal blocks. The Regge limit of a single t-channel conformal block is \cite{Cornalba:2006xm}: 
\bea
g^{0,0,Regge}_{\Delta,j}(\s,\eta)=(2\pi i) \s^{1-j}\eta^{\frac{1}{2}(\Delta-j)}\frac{\Gamma (\Delta +j-1) \Gamma (\Delta +j)}{\Gamma \left(\frac{\Delta +j}{2}\right)^4} {}_{2}F_{1}\bigg(\frac{d-2}{2},\Delta-1,\Delta-\frac{d-2}{2},\eta\bigg). \ 
\eea
We see that the $\s\rightarrow 0$ behavior is now governed by operators with the largest spin. The apparent singular behavior arises because each conformal block has a branch cut, starting at $z=0$ and extending to $-\infty$, which we have crossed.

For CFTs in $d>2$, operators of unbounded spin appear in every OPE and we need to understand how to resum this expansion. This is tractable in large $N$ theories, or theories with a large central charge $C_T$, assuming large $N$ factorization. For such CFTs we can organize the four-point function as follows:
\bea
\<\f\f\psi\psi\>= \<\f\f\>\<\psi\psi\>+\frac{1}{C_T}\<\f\f\psi\psi\>_{c}+ \mathcal{O}\bigg(\frac{1}{C_T^2}\bigg),
\eea

\noindent where we have suppressed higher order terms in $1/C_T$. 

This scaling implies that at large $C_{T}$ the three-point functions behave like: 
\bea
\<\f\psi[\f\psi]_{n,j}\>\sim\ 1, \quad \<\f\f O\>\sim\ \<\f\psi O\>\sim\ \mathcal{O}\big(1/\sqrt{C_T}\big), \quad \<\p\p[\f\f]_{n,j}\>\sim\ \mathcal{O}(1/C_T),
\eea

\noindent where $[\f\psi]_{n,j}\sim \f\partial^{\mu_{1}}...\partial^{\mu_{j}}\partial^{2n}\psi -\mathrm{traces}$ is a double-trace state and $O$ is a single-trace operator. All double-trace states composed of light operators, not necessarily scalars, are required to exist to solve crossing symmetry at order $(C_T)^{0}$, i.e. to match the identity block. 

We can now distinguish between two classes of large $N$ CFTs. The first class occurs when the sum over spin in the $\f\f\rightarrow\p\p$ channel effectively truncates at some $j_{max}$, while the second class occurs when the sum over spins is unbounded. For theories in the former class we can take the Regge limit block-by-block, while for theories in the latter class we need to understand how to resum this expansion at order $1/C_T$. 

Using AdS/CFT it is possible to construct effective CFTs~\cite{Fitzpatrick:2010zm} where the spin effectively truncates at tree-level. A trivial example is when we have a QFT in the bulk consisting of a single $\mathrm{Z}_{2}$ invariant scalar $\phi$ with a finite number of quartic interactions, i.e. $\phi^{4}$, $(\partial \phi)^{4}$, etc. The conformal block decomposition of quartic Witten diagrams is bounded in spin in every channel, so there is no subtlety in going to the Regge limit. A more interesting example will be CFTs dual to gravity plus matter, like $\mathcal{N}=4$ SYM at large $N$ and with the `t Hooft coupling $\lambda \rightarrow \infty$, which we will consider in the next section. In such theories there is a known connection between the large $n,j$ anomalous dimensions of double-trace operators and the phase shift in AdS of a particle propagating through a shockwave \cite{Cornalba:2006xm,Cornalba:2006xk,Cornalba:2007zb}. The precise dictionary for $\<\f\p\p\f\>$, i.e. when $\f$ creates a shock wave which $\p$ traverses, is given by:
\begin{align}
\underset{n,j\rightarrow \infty,\frac{j}{n} fixed}{lim} \gamma_{n,j}=&-\frac{1}{\pi}\delta(s,b), \label{eq:phaseshift}
\\
b=\log\left(\frac{n+j}{n}\right), & \quad s=4n(n+j). \nonumber
\end{align}
Here $\sqrt{s}$ is the total energy of the two-particle state and $b$ is the impact parameter variable.\footnote{In \cite{Camanho:2014apa} they used an alternative AdS parametrization for the impact parameter variable, given by $\rho=\frac{j}{n+j}$.}

This truncation property is not generic and will not hold in $\mathcal{N}=4$ SYM at any finite $\lambda$. To understand solutions to crossing symmetry when an infinite number of spins contribute it will be convenient to use the techniques of conformal Regge theory (CRT), which we will review in section~\ref{sec:CRT}. We will also propose an updated version of the above dictionary for these kinds of theories. 

\section{Bootstrap in the Gravity Limit}
\label{sec:GravityLimit}
In this section we will start by solving the bootstrap equation for CFTs with a large central charge $C_{T}$ in the gravity limit. Our initial goal will be to find the anomalous dimensions and OPE coefficients of double-trace operators in the Regge limit. Concretely, we will focus on the regime $C_T\gg\Delta_{gap}\gg \frac{1}{\sigma}\gg 1$ and expand the bootstrap equation to order $1/C_T$. At infinite $C_{T}$, we will recover the OPE coefficients of the mean field theory. At tree level, we will obtain anomalous dimensions, which exactly match the bulk phase shift computed in \cite{Cornalba:2007zb}. We will also provide a CFT argument that these anomalous dimensions are negative. When generalized to the correlator $\<J \phi \phi J\>$, this condition smoothly interpolates between the conformal collider physics bound and CEMZ constraint for $\<JJT\>$ when we decrease the bulk impact parameter. 

\subsection{Identity Matching}
First let us review the solution of the bootstrap equation at $C_T=\infty$. This solution is simply that of a generalized free theory. We will also take the opportunity to establish some conventions.

For the four-point function of scalars $\langle\phi(0)\psi(z,\bar{z})\psi(1,1)\phi(\infty)\rangle$, we will solve the crossing symmetry equation (\ref{eq:Crossing}).
When $C_{T}$ is infinite the LHS of this sum consists only of the contribution from the identity operator. The bootstrap equation becomes: 
\bea
((1-z)(1-\bar{z}))^{-\Delta_{\p}}(z\bar{z})^{\frac{1}{2}(\Delta_{\f}+\Delta_{\p})}=\sum_{\mathcal{O'}}P_{\mathcal{O}'}g^{a,a}_{\mathcal{O}'}(z,\bar{z}).
\label{eq:LOCrossing}
\eea
On the LHS of this equation, there are power law divergences when $z\rightarrow 1$ and $\bar{z}\rightarrow 1$. They appear because of the OPE singularity when $\psi(z,\bar{z})\rightarrow\psi(1,1)$. On the RHS of this equation, each conformal block only has $\log(1-z)$ and $\log(1-\bar{z})$ singularities. Therefore, to reproduce the leading $\psi\psi$ channel OPE, there must be an infinite number of operators in the $\phi\psi$ channel, which correspond to the familiar double-trace operators $[\phi\psi]_{n,j}$. 

It will be convenient to introduce the variables $h$ and $\bar{h}$, given by
\bea
h=\frac{1}{2}(\Delta+j), \qquad \bar{h}=\frac{1}{2}(\Delta-j).
\eea
In the $\psi\psi\rightarrow \f\f$ OPE limit, the operators that dominate the sum in the $\phi\psi$ OPE are double-trace operators with large spin and large twist. These operators satisfy
\bea
h\approx j +n, \qquad \bar{h}\approx n,
\eea
and, as we will see in more detail shortly, the sum is dominated by the regime
\bea
h\sim \bar{h}\sim \frac{1}{\sqrt{\s}} \label{eq:saddles}.
\eea
So, by taking the limit $\sigma \rightarrow 0$ with $\eta$ finite, we are led to probe the regime $h, \bar{h} \gg 1$ with $\frac{\bar{h}}{h}$ finite \cite{Cornalba:2006xm,Cornalba:2007zb}. 
In this regime, the $\phi\psi$ channel blocks simplify. For example, in 4d, the blocks can be approximated by\footnote{More general cases are described in appendices \ref{App:Conventions} and \ref{App:Impact}.}:
\begin{align}
g^{(d=4),a,a}_{h,\bar{h}}(z,\bar{z})\approx 2^{2(h+\bar{h}-1)}\frac{\sqrt{h\bar{h}}}{\pi}\frac{((1-z)(1-\bar{z}))^{a}}{z-\bar{z}}K_{2a}(2h\sqrt{1-z})K_{2a}(2\bar{h}\sqrt{1-\bar{z}})+(z\leftrightarrow \bar{z}) .  \label{eq:4dBessel}  
\end{align}
Furthermore, in the large $h$ and $\bar{h}$ limit the generalized free theory (or mean field theory) OPE coefficients in general dimensions are given by \cite{Fitzpatrick:2011dm}
\bea
P^{MFT}_{h,\bar{h}}\approx \frac{ 2^{d-2 (h+\bar{h})+2} \pi \left(h^2-\bar{h}^2\right)^{\frac{d}{2}-1} \left(h^{-d+\Delta_{\f}+\Delta_{\p}+\frac{1}{2}} \bar{h}^{-d+\Delta_{\f}+\Delta_{\p}+\frac{1}{2}}\right)}{\Gamma (\Delta_{\f}) \Gamma (\Delta_{\p}) \Gamma \left(-\frac{d}{2}+\Delta_{\f}+1\right) \Gamma \left(-\frac{d}{2}+\Delta_{\p}+1\right)}, \label{eqn:MFT}
\eea
and we can approximate the $\phi\psi$ channel sum as an integral over $h$ and $\bar{h}$. With the approximate conformal blocks above, the basic integral we need is then
\be
\int_{0}^{\infty}dh h^p K(2h\sqrt{1-z})\propto \frac{1}{(1-z)^{\frac{1}{2}(p+1)}}.
\ee
Applying these approximations to the RHS of (\ref{eq:LOCrossing}) and restricting to the wedge $h \geq \bar{h}$, we can readily reproduce the large $h$, $\bar{h}$ limit of the generalized free theory OPE coefficients, given in (\ref{eqn:MFT}), for $d=4$. The precise integrals needed to do this matching are provided in appendix~\ref{App:IntK}. 

This procedure is similar to the lightcone bootstrap initiated in \cite{Fitzpatrick:2012yx,Komargodski:2012ek}, but here we are just matching the standard $\p\p\rightarrow\f\f$ OPE limit. The $\phi\psi$ channel sum must reproduce the power law singularity in both $1-z$ and $1-\bar{z}$. So instead of a single sum over spin, or $h$, we need two infinite sums over both $h$ and $\bar{h}$. 

\subsection{Single-Trace Matching}
\label{subsec:GravityNLOMatching}
Our next step is to solve the bootstrap equation at leading order in $1/C_T$ for the exchange of an isolated single-trace operator in the t-channel. If we stay in the OPE limit, it is not immediately straightforward how to match this exchange to an infinite sum in the s-channel. In part this is because one must disentangle corrections to double-trace contributions from other new operators that appear at order $1/C_T$.  Instead, we will move to the Regge limit, where we have analytically continued $z\rightarrow e^{-2\pi i}z$. This regime has the advantage that single-trace contributions in the s-channel are suppressed relative to double-trace contributions, allowing us to more cleanly match the bootstrap equations.

The $\psi\psi$ channel (t-channel) may contain various single-trace and double-trace operators at order $1/C_T$. In $d=4$, the contribution of each operator as $\sigma \rightarrow 0$ is 
\bea
g^{Regge}_{\Delta,j}(\s,\eta)=(2\pi i) \s^{1-j}\frac{\eta^{\frac{1}{2}(\Delta-j)}}{1-\eta}\frac{\Gamma (\Delta +j-1) \Gamma (\Delta +j)}{\Gamma \left(\frac{\Delta +j}{2}\right)^4}.
\label{eq:ScalarSpinjGeneralD}
\eea
We will show soon that when constructing the minimal solution to crossing to reproduce a single-trace contribution in the t-channel, we automatically produce double-trace contributions in the t-channel as well. The full answer reproduces exactly the Regge limit of an exchange Witten diagram. In appendix~\ref{App:DoubleTrace}, we study the full effect of these double-trace states for a specific correlator. 

Considering isolated single-trace exchange is also motivated by the gravity limit of holographic theories where the spin of the single-trace spectrum is bounded. Then the chaos bound~\cite{Maldacena:2015waa} implies that $j\le2$. Another possibility is having towers of operators with unbounded spin that resums into a softer effective spin $j(0)\le2$ in the Regge limit, which will be considered in section~\ref{sec:ReggeExchange}.

Next, we consider expanding the four-point function in the $\phi\psi$ channel (s-channel). Each conformal block on the first sheet has the following small $z,\bar{z}$ expansion:
\bea
g^{a,b}_{\mathcal{O}'}(z,\bar{z})=\sum_{n,m}a_{h,\bar{h}}z^{\frac{1}{2}(\Delta-j)+n}\bar{z}^{\frac{1}{2}(\Delta+j)+m}+(z\leftrightarrow \bar{z}),
\eea
where $n$ and $m$ are integers, representing the sum over all descendants. Taking $z\rightarrow e^{-2\pi i}z$ to go to the second sheet, the block picks up a phase factor proportional to its twist: 
\bea
g^{a,b}_{\mathcal{O}'}(z,\bar{z})\rightarrow e^{-i\pi(\Delta'-j')}g^{a,b}_{\mathcal{O}'}(z,\bar{z}).
\eea
The most general bootstrap equation at leading order in $1/C_T$ in the Regge limit is then:
\begin{small}
\begin{align}
\hspace{-.5cm}
e^{-i \pi (\Delta_{\f}+\Delta_{\p})}&\s^{-2\Delta_{\p}}\eta^{-\Delta_{\p}} \mathcal{A}(\s,\eta) \nonumber\\
=\,& e^{-i\pi(\Delta_{\f}+\Delta_{\p})}\sum_{n,j}P^{MFT}_{h,\bar{h}}\left[ \gamma_{h,\bar{h}}\big(-i\pi+\frac{1}{2}(\partial_{h}+\partial_{\bar{h}})\big)+ \delta P_{h,\bar{h}}\right]g^{a,a}_{h,\bar{h}}(1-\s,1-\eta \s) \nonumber \\ &+\sum_{\mathcal{O'}}e^{-i\pi\tau_{\mathcal{O}'}}P_{\mathcal{O'}}g^{a,a}_{\mathcal{O}'}(1-\s,1-\eta \s),
\label{eq:NLOCrossing}
\end{align}
\end{small}
where $\mathcal{A}(\s,\eta)$ denotes the leading Regge contribution to the four-point function computed in the $\psi\psi\rightarrow \f\f$ channel. The first sum on the right hand side runs over the double-trace states $[\f\p]_{n,j}$, where we consider $1/C_T$ corrections to the dimensions and OPE coefficients, denoted by $\gamma_{h,\bar{h}}$ and $\delta P_{h,\bar{h}}$. The second sum runs over operators which first appear at order $1/C_T$. It is important to note that the double-trace operators always add in phase while the new operators, labelled as $\mathcal{O}'$, will generically not add in phase. Therefore, typically their contributions are small in the Regge limit compared to the double-trace ones, and going forward we will assume that there are no additional towers of operators that add in phase. 

Thus, to match an isolated, single-trace operator in the $\psi\psi\rightarrow\f\f$ channel, i.e. a single Regge block (\ref{eq:ScalarSpinjGeneralD}), we only need to use the anomalous dimensions in (\ref{eq:NLOCrossing}). This is because, for $\s$ real, the Regge block is purely imaginary and only the anomalous dimensions contribute to the imaginary part of the four-point function. In particular, the anomalous dimensions that match a Regge block in $d=4$ are:  
\bea
\gamma_{h,\bar{h}}&=&-\gamma_{0}C_{\f\f\mathcal{O}}C_{\p\p\mathcal{O}}(h\bar{h})^{j-1} \left(\frac{\bar{h}}{h}\right)^{\Delta-1} \frac{h^{2}}{h^2-\bar{h}^2}, \label{eqn:anomST}
\\
\gamma_{0}&=&\frac{2\left(\frac{1}{2}\right)^{j}\Gamma (\Delta_{\f}-1) \Gamma (\Delta_{\f}) \Gamma (\Delta_{\p}-1) \Gamma (\Delta_{\p})  \Gamma (\Delta +j-1) \Gamma (\Delta +j)}{\Gamma \left(\frac{\Delta +j}{2}\right)^4 \Gamma \left(-\frac{\Delta }{2}+\Delta_{\f}+\frac{j}{2}\right) \Gamma \left(\frac{1}{2} (\Delta +2 \Delta_{\f}+j-4)\right) \Gamma \left(-\frac{\Delta }{2}+\Delta_{\p}+\frac{j}{2}\right) \Gamma \left(\frac{1}{2} (\Delta +2 \Delta_{\p}+j-4)\right)}. \nonumber \\
\eea
To derive this, we approximate the sum over $j$ and $n$ as integrals over $h$ and $\bar{h}$ and match all terms of the form $\sigma^{1-j} \eta^{\frac12(\Delta-j) +n}$, where $n$ is an integer, after expanding at small $\sigma$ and $\eta$. Note that by selecting this power of $\eta$, we are effectively mapping out double-trace contributions in the $\psi\psi\rightarrow\f\f$ channel.\footnote{For isolated values of the external dimensions $\Delta_{\psi,\phi}$, the double-trace states may also contribute to $\eta^{\frac12(\Delta-j) +n}$ type terms in the amplitude. Our selection procedure is unambiguous for generic scaling dimensions and we can obtain results for these special cases using continuity and the uniqueness of the solution \cite{Heemskerk:2009pn}.} 
As described in appendix~\ref{App:IntK}, this is done using the following master integral:
\bea
\int_{0}^{\infty} d\bar{h} \int_{\bar{h}}^{\infty} dh \ 2^{-2(h+\bar{h})}h^{c_{1}}\bar{h}^{c_{2}}g^{a,a}_{h,\bar{h}}(1-\s,1-\eta\s)\sim \s^{\frac{1}{2} (-4 a-c_{1}-c_{2}-5)}\frac{\eta^{\frac{1}{4} (-4 a-2 c_{1}-3)}}{1-\eta} + \ldots,
\label{eq:MasterIntegral}
\eea 
where the ellipses denote terms with different powers of $\eta$ which will correspond to double-trace contributions in the t-channel. 

The large $h,\bar{h}$ behavior of the anomalous dimension $\gamma_{h,\bar{h}} \sim (h\bar{h})^{j-1}$ is controlled by the spin $j$ of the exchanged single-trace operators in the $\psi\psi$ channel. Since the AdS-dual bulk scattering energy is given by $s= 4 h \bar{h}$, this is the familiar Regge limit behavior. The precise form of the anomalous dimensions also agrees with previous results in $d=4$ for AdS graviton exchange~\cite{Cornalba:2006xm,Cornalba:2006xk,Cornalba:2007zb}. We can see this exact agreement by focusing on CFT stress tensor exchange, $\Delta=4$ and $j=2$, and using the general relations:
\bea
G_{N}^{(5)}=\frac{20\pi}{C_{T}}, \qquad C_{\f\f T}= -\frac{4\Delta_{\f}}{3 \sqrt{C_{T}}}.
\eea
The anomalous dimensions matching $d=4$ stress tensor exchange are then:
\be
\gamma_{h,\bar{h}}=- \frac{8G_N^{(5)}}{\pi}\frac{\bar{h}^4}{h^2-\bar{h}^2} =  -\frac{160}{C_T} \frac{\bar{h}^4}{h^2-\bar{h}^2}. \label{eq:adgravity}
\ee
Note that the external dimensions $\Delta_{\phi,\psi}$ drop out of this result. This is intuitive since bulk gravity is sourced by the energy of the colliding particles and we are working in the regime $h,\bar{h}\gg\Delta_{\phi,\psi}$, where this energy is dominated by its kinetic part, rather than the rest mass of the particles. One can also repeat this exercise in $d=2$ and see exact agreement. For a discussion of how to do this matching in general dimensions see appendix \ref{App:Impact}.

As emphasized above, we have been focusing on matching the single-trace operator in the $\psi\psi$ channel by looking at terms proportional to $\sigma^{1-j} \eta^{\frac{1}{2}(\Delta-j)+n}$, where $n$ is a positive integer. However, if we plug our solution (\ref{eqn:anomST}) into (\ref{eq:NLOCrossing}) (using the integrals in appendix~\ref{App:IntK}), we also obtain terms of the form $\s^{1-j}\eta^{\Delta_{\f}+p}$ and $\s^{1-j}\eta^{\Delta_{\p}+q}$, for $p$ and $q$ positive integers. These are precisely the right powers to correspond to double-trace states, $[\f\f]_{n,j}$ and $[\p\p]_{n,j}$ respectively, appearing in the $\psi\psi$ channel. These double-trace contributions correctly dress the single-trace conformal block into a bulk Witten diagram. We work out specific examples to demonstrate this fact in appendix~\ref{App:DoubleTrace}. We will also derive this more elegantly in section \ref{sec:ReggeExchange}. 

Although we kept $j$ general, in theories where the t-channel contains operators with bounded spin at order $1/C_T$, all $j\ge3$ operators will be forbidden by the chaos bound. The leading contribution to $\gamma_{h,\bar{h}}$ then comes from the stress tensor $T_{\mu\nu}$, which is guaranteed to appear by a conformal Ward identity, and possibly other spin-2 single-trace operators, $\mathcal{O}_{\Delta,j=2}$. These operators have necessarily larger dimension $\Delta$, so in the limit $\bar{h}\ll h$ these operators will be suppressed like $(\bar{h}/h)^{\Delta}$. For generic values of $\bar{h}/h$ though they contribute at the same order as the stress tensor.

\subsection{Current-Scalar Correlators}
\label{subsec:Current-scalars-T}
We will now generalize our discussion to a correlation function of conserved currents $J^{\mu}$ and scalars $\f$:
\bea
G_J(z,\bar{z},\{\epsilon_{i}\})=\<\epsilon_{1}\cdot J(0)\f(z,\bar{z})\f(1) \epsilon_{4}\cdot J(\infty)\>,
\eea
with the current normalized as
\bea
\<J^{\mu}(x)J_{\nu}(0)\>=C_{J}\frac{\delta^{\mu}_{\nu}-\frac{2x^{\mu}x_{\nu}}{x^{2}}}{x^{2\Delta_{J}}}.
\eea
In general dimensions the conservation condition implies $\Delta_{J}=d-1$. 

The two conformal block decompositions are: 
\begin{align}
\text{$J\phi$-channel: \hspace{.25cm}} G_J(z,\bar{z},\{\epsilon_{i}\}) &=(z\bar{z})^{-\frac{1}{2}(\Delta_{\f}-\Delta_{J})}\sum_{\mathcal{O}}\left(-\frac12\right)^j C_{J\phi\mathcal{O}}C_{\phi J\mathcal{O}} g^{a,a}_{p,\mathcal{O}}(z,\bar{z}) Q^{(p)}(z,\bar{z},\{\epsilon_{i}\})\label{eq:JJOOsch} \\ \nonumber
\hspace{1.6cm}&= G_{J,[k]}(z,\bar{z},\{\epsilon_{i}\})+ G_{J,[k+1,1]}(z,\bar{z},\{\epsilon_{i}\}),
\\
\text{$JJ$-channel:  \hspace{.25cm}} G_J(z,\bar{z},\{\epsilon_{i}\})&=[(1-z)(1-\bar{z})]^{-\Delta_{\f}}\sum_{\mathcal{O},b} \left(\frac12\right)^j C^b_{JJ\mathcal{O}}C_{\f \f \mathcal{O}}g^{0,0}_{\mathcal{O},b,p}(1-z,1-\bar{z})Q^{(p)}(z,\bar{z},\{\epsilon_{i}\})  \label{eq:JJOOtch},
\end{align}
with $a=-\frac{1}{2}(\Delta_\f-\Delta_J)$, and in the s-channel we have explicitly separated the sum over the symmetric traceless tensors (labelled by $[k]$) and the mixed symmetry operators (labelled by $[k+1,1]$). We will once again define:
\bea
P_{\mathcal{O}}=\left(-\frac12\right)^j C_{J\phi\mathcal{O}}C_{\phi J\mathcal{O}}=\left(\frac12\right)^j (C_{J\phi\mathcal{O}})^{2}
\eea
to simplify the notation. Schematically the double-trace operators in these representations take the form
\be
[J\f]_{[k],n}\approx J^{\mu_1} \partial^{(\mu_2}\dots\partial^{\mu_{k})}\partial^{2n}\phi, \hspace{1cm}[J\f]_{[k+1,1],n}\approx J^{[\mu_1} \partial^{(\mu_2]}\partial^{\mu_3}\dots\partial^{\mu_{k+2})}\partial^{2n}\phi.
\label{eq:JPhiDoubleTraces}
\ee
The label $p$ indexes different linearly independent tensor structures and the label $b$ in the t-channel decomposition distinguishes the multiple possible three-point function structures in $\<JJ\mathcal{O}\>$. The three-point functions for $\<J\f\mathcal{O}\>$ are unique once we impose conservation. Moreover, in the t-channel only symmetric traceless operators of even spin can appear because we are also considering the OPE of identical scalars.

Adopting the notation of \cite{Rejon-Barrera:2015bpa}, the tensor structures are given by:
\bea
Q(z,\bar{z},\{\epsilon_{i}\})=\{m^{(14)},k^{(123)}k^{(413)},k^{(123)}k^{(423)},k^{(134)}k^{(413)},k^{(134)}k^{(413)}\},
\\
m^{(ij)}=\epsilon_{i}\cdot\epsilon_{j}-\frac{2}{x_{ij}^{2}}\epsilon_{i}\cdot x_{ij}\epsilon_{j}\cdot x_{ij}, \qquad k^{(ijk)}=\frac{x_{ij}^{2}\epsilon_{i}\cdot x_{ik}-x_{ik}^{2}\epsilon_{i}\cdot x_{ij}}{(x_{ij}^{2}x_{ik}^{2}x_{jk}^{2})^{\frac{1}{2}}}.
\eea 
Conformal blocks for the $[k]$ representations can be calculated using the differential operator method of \cite{Costa:2011dw}, while for the $[k+1,1]$ blocks we will use the results of \cite{Rejon-Barrera:2015bpa}. To simplify the calculations, we perform the $h$ and $\bar{h}$ integrals first and then apply the differential operations needed to obtain the spinning conformal blocks. 

We proceed in a similar way as in the last section. We first solve the generalized free theory bootstrap equations by matching the identity operator in the $JJ\rightarrow\f\f$ channel with double-trace operators in the $J\phi$ channel. Solving this equation provides the generalized free theory OPE coefficients of the double-trace operators in the large $h$ and $\bar{h}$ regime. Unlike the scalar case, as far as we are aware these coefficients were not known before. We find 
\bea
P^{MFT}_{[k],h,\bar{h}}=C_{J}\frac{16\pi(h^{2}-\bar{h}^{2})(h\bar{h})^{\Delta_{\f}-\frac{3}{2}}}{3\Gamma(\Delta_{\f}-1)\Gamma(\Delta_{\f})},
\\
P^{MFT}_{[k+1,1],h,\bar{h}}=C_{J}\frac{512\pi(h^{2}-\bar{h}^{2})(h\bar{h})^{\Delta_{\f}-\frac{1}{2}}}{3\Gamma(\Delta_{\f}-1)\Gamma(\Delta_{\f})}.
\eea

At leading order, we again focus on the single-trace contributions in the $JJ$ channel and will consider stress tensor exchange in the $JJ$ channel. The three-point function $\<JJT\>$ can be given the parametrization
\bea
\<JJT\>=\<JJT\>_{\textrm{Maxwell}}+a_{2}\<JJT\>_{\textrm{Weyl}} .
\eea 
The Maxwell structure is generated by a tree-level $F^{2}$ term in the bulk dual Lagrangian while the Weyl structure is generated by a $W^{\mu\nu\delta\rho}F_{\mu\nu}F_{\delta\rho}$ term, where $W$ is the Weyl tensor. Then $a_{2}$ is a three-point function coefficient which will measure the deviation of the dual theory from Einstein-Maxwell theory, at the level of the cubic couplings \cite{Hofman:2008ar}.

The stress tensor contribution is then matched by the anomalous dimensions of the double-trace operators in the $J\phi$ channel, which we find to be
\bea
\gamma_{[k],h,\bar{h}}^{(T)}=40\bar{h}^{4}C_{\f\f T}\frac{2a_{2}h^{4}+3(h^2-\bar{h}^2)^2}{C_{T}\Delta_{\f}(h^{2}-\bar{h}^{2})^{3}}, \label{eq:adSTT}
\\
\gamma_{[k+1,1],h,\bar{h}}^{(T)}=40\bar{h}^{4}C_{\f\f T}\frac{-a_{2}h^{4}+3(h^2-\bar{h}^2)^2}{C_{T}\Delta_{\f}(h^{2}-\bar{h}^{2})^{3}}. \label{eq:adMixed}
\eea

\begin{figure}
\begin{centering}
\includegraphics[scale=0.5]{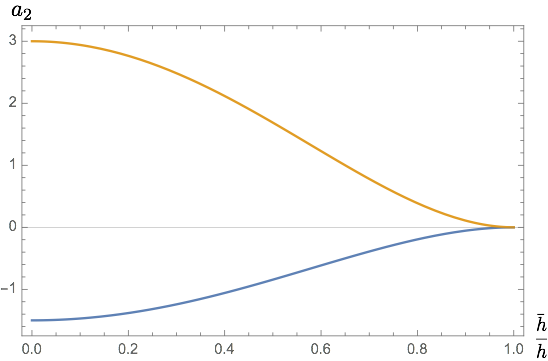}
\par\end{centering}

\caption{The bounds on the coefficient $a_{2}$ of $\langle J_{\mu}J_{\nu}T_{\rho\sigma}\rangle$
derived from the negativity of the anomalous dimensions in the Regge limit $\gamma^{(T)}_{h,\bar{h}}\le0$
as a function of $\bar{h}/h$. 
}
\end{figure}

If we require these anomalous dimensions to be negative for all values of $h$ and $\bar{h}$, we obtain the bound
\bea
-\frac{3}{2}\left(\bar{h}^{2}/h^{2}-1\right)^2\leq a_{2}\leq 3 \left(\bar{h}^{2}/h^{2}-1\right)^2 . \label{eq:a2bound}
\eea
When $\bar{h}\ll h$ we recover the conformal collider bounds, $-\frac{3}{2}\leq a_{2}\leq 3$, while taking $\bar{h}\rightarrow h$ we recover the CEMZ bound, $a_{2}=0$ \cite{Camanho:2014apa}. Using (\ref{eq:phaseshift}), the anomalous dimensions match exactly the two possible time delays a gauge boson can experience when crossing a gravitational shockwave \cite{Camanho:2014apa}. 

It is interesting to ask how this bound is modified if other single-trace operators are included. The contribution of operators $\mathcal{O}_{\Delta,j}$ of dimension $\Delta_{\cO}$ and spin $j$ to the anomalous dimensions $\gamma_{h,\bar{h}}$ has the form:
\bea
\gamma_{h,\bar{h}}\propto(h\bar{h})^{j-1}(\bar{h}/h)^{\Delta_{\cO}-1}(1+...),
\eea
so if the operators are very heavy, $\Delta_{\mathcal{O}}\gg1$, then their contribution becomes significant when $1-\bar{h}/h\sim\Delta_{\mathcal{O}}^{-1}$. We must also assume $j\geq2$ so they are not sub-leading in the Regge limit. Plugging this approximation into (\ref{eq:a2bound}) yields 
\begin{align*}
|a_{2}|\lesssim \frac{1}{\Delta_{\mathcal{O}}^{2}},
\end{align*}
where $\mathcal{O}$ is the lightest operator with spin $j\geq 2$ beyond the stress tensor.

In general we will not be able to say that requiring negative anomalous dimensions requires $|a_{2}|\lesssim \frac{1}{\Delta_{gap}^{2}}$, where $\Delta_{gap}$ is the gap between the stress tensor and the lightest operator with spin $j\geq4$. Other spin-2 single-trace operators may also contribute to the anomalous dimensions and we would need to include their affect on the anomalous dimensions. Therefore, the bound on $a_{2}$ we derive will be sensitive to both the spin-2 and higher spin sector of the theory.

\subsection{Negativity of Anomalous Dimensions}
\label{subsec:Holographic-Bound}
In this section we will give an argument that the chaos bound implies that the anomalous dimensions of double-trace operators of large spin and large twist are negative $\gamma_{h,\bar{h}}\le 0$. For the time being we will assume that spin-2 operators dominate the t-channel Regge limit, so that the dimension $\Delta_{gap}$ at which higher spin single-trace operators appear is sent to infinity. As in the previous section we will be in the regime $C_T\gg\Delta_{gap}\gg h\sim\bar{h}\gg1$.

We will first give an argument for a scalar four-point function. We then comment on additional features when operators have spin. Let us return to the four-point function
\bea
G(z,\bar{z})=\<\psi\phi\phi \psi\> .
\eea
Reflection positivity guarantees that in the $\psi\phi$ channel decomposition, the coefficients are positive \cite{Fitzpatrick:2012yx,Hartman:2015lfa}:
\bea
G(z,\bar{z})=\sum_{h,\bar{h}}a_{h,\bar{h}}z^{h}\bar{z}^{\bar{h}}, \qquad a_{h,\bar{h}}\geq0.
\label{eq:PositiveCoeff}
\eea
The positivity $a_{h,\bar{h}}\ge0$ implies that when we continue to the second sheet, $z\rightarrow ze^{-2\pi i}$, each term will produce a phase and the sum is bounded in terms of the first sheet correlator. Therefore, $\widehat{G}(z,\bar{z}) \equiv G(z e^{-2\pi i},\bar{z})$ is bounded in terms of $G(z,\bar{z})$ and one can show $\widehat{G}$ is analytic in the complex $\s$ plane, minus the point $\s=0$ corresponding to a Euclidean OPE singularity \cite{Hartman:2015lfa}. 

The spin-2 contribution on the LHS leads to the following form of $\widehat{G}$ in the Regge limit:
\bea
\frac{\widehat{G}(z,\bar{z})}{\<\phi \phi\>\<\psi \psi\>}=1+i \frac{1}{C_T}  \frac{f(\eta)}{\s},
\hspace{1cm}C_T\gg\Delta_{gap}\gg\sigma^{-1}\gg1.
\label{eq:ChaosBoundCorrelator}
\eea
Then the chaos bound implies that $f(\eta)>0$.\footnote{The sign flip in comparison to \cite{Hartman:2015lfa} is because $\s_{there}=-\s_{here}$.} In the $\psi\phi$ channel decomposition, $f(\eta)$ is given by a sum over double-trace anomalous dimensions:
\bea
f(\eta)=-\pi \s \sum_{h,\bar{h}}P^{MFT}_{h,\bar{h}}\gamma_{h,\bar{h}}g^{a,a}_{h,\bar{h}}(1-\s,1-\eta \s).
\eea
At this point, we note that crossing symmetry already determines the $h$ and $\bar{h}$ dependence, as seen in e.g. (\ref{eqn:anomST}). The bound above then fixes the sign the anomalous dimensions to be negative for all $\Delta_{gap}\gg h\gtrsim \bar{h}\gg1$. 
We sketch an alternative argument requiring partial wave positivity in appendix~\ref{app:PartialWavePositivity}.

One can try to make the same argument for four-point functions involving spinning operators
\bea
G^{(j_{1},j_{2})}(z,\bar{z},\{\epsilon_{i}\})=\<\epsilon_{1}\cdot\mathcal{O}_{1}\epsilon_{2}\cdot\mathcal{O}_{2}\epsilon_{3}\cdot\mathcal{O}_{2}\epsilon_{4}\cdot\mathcal{O}_{1}\>.
\eea

There are a few additional features. One is that we need to choose the external polarizations in a reflection positive way, in the context of radial quantization. Denoting $\epsilon^{R}=\mathcal{I}\cdot \epsilon$ where $\mathcal{I}$ is the inversion tensor in the appropriate representation, we need $\epsilon_{4}=\epsilon_{1}^{* R}$ and $\epsilon_{3}=\epsilon_{2}^{* R}$ to have reflection positivity. We also need to restrict the decomposition (\ref{eq:sumrepRestriction}) further by projecting onto operators in a given Lorentz representation. For example, for $\<J\f\f J\>$ we need to separately project onto the families of double-trace operators in (\ref{eq:JPhiDoubleTraces}). The resulting negativity properties needs to hold for each family. 
The argument for this requires that each projected partial wave sum define an analytic function in an appropriate region of the complex $\sigma$ plane which is bounded along the real $\sigma$ line in both the $s$-channel regime $\sigma > 0$ and the $u$-channel regime $\sigma < 0$. While this can be explicitly checked in cases where the double-trace operators dominate the sum, we have not yet understood a rigorous argument establishing this property in the $u$-channel regime when $|\sigma| \lesssim 1/C_T$.\footnote{We thank Sasha Zhiboedov and Tom Hartman for discussions on this issue.} We hope this can be done in future work, in particular by focusing on the double discontinuity of the projected sum. Assuming this property holds, this result, together with the relation between the phase shift of bulk scattering and the anomalous dimensions (\ref{eq:phaseshift}), would show that in correlators with spinning operators, AdS causality is also a consequence of unitarity of the boundary CFT.

Applying this bound for $\<J\f\f J\>$, assuming there are no single-trace spin-2 operators besides the stress tensor when we take $\Delta_{gap}\rightarrow\infty$, implies that we must have $a_{2}=0$. Given the close connection between anomalous dimensions and causality (\ref{eq:phaseshift}) \cite{Cornalba:2007zb,Camanho:2014apa}, we expect that when this bound is applied for $\<TTTT\>$ one will recover the full CEMZ bound. 

\section{Beyond the Gravity Limit}
\label{sec:ReggeExchange}
As we probe bulk distances that are sufficiently small, or equivalently go to sufficiently high energies, we have to consider the non-local nature of the underlying theory of quantum gravity in AdS.

There are two sources of non-locality. One is quantum and the other one is classical. If the energy of the collision $s\sim \frac{1}{\sigma}\sim C_T$, then the large $N$ perturbative expansion is no longer valid. Generically there could be events such as black hole creation that prevent locality to hold at such small distances. Locality can also break down classically when $s\sim \frac{1}{\sigma}\sim\Delta_{gap}$. Then an infinite tower of higher spin states become important and one needs to go beyond graviton exchange. In string theory this type of non-locality arises from the extended nature of strings, but our CFT based discussion will be more general.

In this paper, we work to leading order in $C_T^{-1}$, so we will not probe the quantum break down of locality. However, by increasing the collision energy we are fully equipped to understand the classical non-locality, if there is a scale separation for it to appear well below the Planck scale. Therefore, we focus on CFTs with $\Delta_{gap} \ll C_T$. In particular, we will consider the effect of the entire leading Regge trajectory in the $\psi\psi\rightarrow \f\f$ channel. Then we will discuss various examples, such as when $\Delta_{gap}$ is large but finite. 

In the $\psi\psi$ channel, we use the techniques of conformal Regge theory developed in \cite{Costa:2012cb} to compute the contribution of a generic leading Regge trajectory. We then solve the bootstrap equation and obtain anomalous dimensions and leading corrections to the OPE coefficients. We apply the chaos bound to the four-point function in this regime and obtain generic constraints on the Regge trajectory. In the $\psi\phi$ channel, the chaos bound further implies the negativity of anomalous dimensions and corrections to the OPE coefficients when $1\le j(0)\le2$, which implies AdS causality and unitarity. 

In addition, we show that to complete these results into a physical solution of the crossing equation, there must exist an infinite number of new single-trace operators in the $\psi\phi$ OPE. In a bulk string theory, these operators should correspond to massive string states created in the high-energy collision of light states.

\subsection{The Leading Regge Trajectory}
\label{sec:CRT}
In this section we will briefly review the results and notation of \cite{Costa:2012cb}. Our conventions will differ slightly because we take $z\rightarrow ze^{-2\pi i}$. Assuming single Regge pole dominance, they find that resuming the contribution of the leading Regge trajectory to the four-point function gives
\bea
\mathcal{A}(\s,\eta)\approx 2\pi i \int_{-\infty}^{\infty}d\nu\alpha(\nu)(\s\sqrt{\eta})^{1-j(\nu)}\Omega_{i\nu}\left(-\frac{1}{2}\log(\eta)\right). \label{eq:4ptRegge}
\eea

To unpack this formula we start with $j(\nu)$, an even function of $\nu$, which is related to the spectrum of the leading Regge trajectory by:
\bea
\nu^{2}+(\Delta(j(\nu))-d/2)^{2}=0.
\eea
In other words, the scaling dimension $\Delta$ and the parameter $\nu$ are related by $\Delta=\frac{d}{2}+i\nu$, and $\Delta(j)$ gives the physical spectrum of the leading trajectory. So, physical dimensions correspond to imaginary values of $\nu$. We will assume $j(\nu)$ is regular around $\nu=0$. We also assume the existence of the stress tensor, fixing $j(\pm2i)=2$. As derived in~\cite{Costa:2012cb}, the function $\alpha(\nu)$ is determined by an analytic continuation of the OPE coefficients and the function $j(\nu)$ via the relation
\bea
\alpha(\nu)=\frac{\pi^{h-1}2^{j(\nu)-1}e^{\frac{-i\pi j(\nu)}{2}}}{\sin(\pi j(\nu)/2)}\gamma(\nu)\gamma(-\nu)\frac{\pi}{4\nu} j'(\nu) K_{\Delta(j(\nu)),j(\nu)} C_{\f\f j(\nu)} C_{\p\p j(\nu)}. \label{eq:alpha}
\eea 

When $j(\nu_{*})=\ell$ is an even integer, $C_{\f\f j(\nu_{*})}=C_{\f\f\mathcal{O}_{\Delta,\ell}}$ for $\mathcal{O}_{\Delta,\ell}$ on the leading Regge trajectory. Since the resummation is done in terms of the Mellin amplitude partial waves, we can think of this as a Regge resummation in terms of Witten diagrams, as opposed to in terms of individual conformal blocks. In a purely CFT language, this means we are doing a resummation in terms of solutions to crossing symmetry. This fact is encoded in the function $\gamma(\nu)$:
\bea
\gamma(\nu)=\Gamma\left(\frac{1}{2}(2\Delta_\f+j(\nu)+i\nu-\frac{d}{2})\right)\Gamma\left(\frac{1}{2}(2\Delta_\p+j(\nu)+i\nu-\frac{d}{2})\right) .
\label{eq:gammaNu}
\eea

If we set $j(\nu)=2$, i.e. if we want to consider a theory with no higher spin, single-trace states, and close the $\nu$ contour in the upper half plane, then the poles of $\gamma(\nu)$ encode the contributions of $[\f\f]_{n,j=2}$ and $[\p\p]_{n,j=2}$, which appear in the direct channel decomposition of a spin-2 Witten diagram.\footnote{When fixing $j(\nu)$ to be an integer one has to be careful cancelling zeros in $\alpha(\nu)$.} In general, the Regge limit of a spin-$j$ Witten diagram determines the couplings $\<\p\p[\f\f]_{n,j}\>$ and $\<\f\f[\p\p]_{n,j}\>$ for the double-trace states with maximal spin $j$ in the direct channel.

Finally, the function $\Omega_{i\nu}(\rho)$ is a harmonic function on $H_{d-1}$, or $d-1$ dimensional hyperbolic space. For $d=4$ it is given by:
\bea
\Omega_{i\nu}(\rho)=\frac{\nu \sin(\rho \nu)}{4\pi^{2}\sinh(\rho)} .
\label{eq:4dReggeBlock}
\eea 

\noindent For the explicit form of $K_{\Delta,J}$ and $\Omega_{i\nu}(\rho)$ in arbitrary dimensions see appendix \ref{App:Conventions}. 

\subsection{Crossing Symmetry}
\label{subsec:ReggeTrajectoryCrossing}
In this section, we study the matching of the entire Regge trajectory in the crossed channel. 
We compute the anomalous dimensions and corrections to the OPE coefficients in the $\psi\phi$ channel that match to the leading Regge trajectory in the $\psi\psi$ channel.
Rather than match to (\ref{eq:saddle}) directly it is simpler to match to (\ref{eq:4ptRegge}) under the $\nu$ integral. We assume we can write $\gamma_{h,\bar{h}}=\int d\nu \gamma_{h,\bar{h}}(\nu)$ and $\delta P_{h,\bar{h}}=\int d\nu \delta P_{h,\bar{h}}(\nu)$. 

To keep the presentation simpler we will assume $\s$ is real when doing the matching. Then matching the anomalous dimensions and corrections to the OPE coefficients corresponds to matching the imaginary and real parts of the correlator, respectively. We find the following equation for the $\gamma_{h,\bar{h}}$:
\be
\Re (\sigma^2 \eta)^{-\De_2}  \sigma^{1-j(\nu)} \frac{\eta^{1-j(\nu)/2}}{1-\eta} \frac{\nu \frac{i}{2} ( \eta^{i\nu/2} - \eta^{-i \nu/2})}{2\pi^2 } \alpha(\nu) &=& - \int d h d \bar{h} P_{h,\bar{h}}^{MFT} \frac{1}{2}\gamma_{h,\bar{h}}(\nu) g_{h,\bar{h}}(z,\bar{z}),\nn\\
\ee

\noindent where we replaced $\sin(\nu \log(1/\sqrt{\eta})) = \frac{i}{2} ( \eta^{i\nu/2} - \eta^{-i \nu/2})$ and $\sinh(\log(1/\sqrt{\eta})) = \frac{1-\eta}{2\sqrt{\eta}}$. To simplify the integrals we will use an ansatz that $\gamma_{h,\bar{h}}(\nu)$ is symmetric in $h$ and $\bar{h}$. This symmetry property of the integrand does not necessarily extend to the full integral, i.e. $\gamma_{h,\bar{h}}$, as we will demonstrate below. To see why this property is useful, one can note that in $d=4$ we have $P^{MFT}_{h,\bar{h}}= -P^{MFT}_{\bar{h},h}$ and $g^{a,a}_{h,\bar{h}}(z,\bar{z})= -g^{a,a}_{\bar{h},h}(z,\bar{z})$. Therefore if $\gamma_{h,\bar{h}}(\nu)=\gamma_{\bar{h},h}(\nu)$ we can write, restoring the integration bounds,
\bea
\int_{0}^{\infty} d h \int_{0}^{h}d \bar{h} P_{h,\bar{h}}^{MFT} \frac{1}{2}\gamma_{h,\bar{h}}(\nu) g_{h,\bar{h}}(z,\bar{z})=\int_{0}^{\infty} d h \int_{0}^{\infty}d \bar{h} P_{h,\bar{h}}^{MFT} \frac{1}{4}\gamma_{h,\bar{h}}(\nu) g^{a,a}_{h,\bar{h}}(z,\bar{z}).
\eea 

With this assumption and the explicit form of the 4d blocks, the $h$ and $\bar{h}$ integrals factorize, and we only need the following simple integral:
\bea
\int_{0}^{\infty} dh\ h^{a}K_{b}(2h\sqrt{z})=\frac{1}{4} z^{-\frac{a}{2}-\frac{1}{2}} \Gamma \left(\frac{1}{2} (a-b+1)\right) \Gamma \left(\frac{1}{2} (a+b+1)\right).
\eea

We can then use this integral to obtain
\be
\gamma_{h,\bar{h}}(\nu) &=& -\Re \left( \frac{\nu \alpha(\nu)}{\pi^2} \right) \frac{\gamma_{0}}{h^2-\bar{h}^2}  \left[ \frac{i}{2}\left( h^{a^-} \bar{h}^{a^+} - h^{a^+} \bar{h}^{a^-}\right) \right],
\ee
with 
\be
a^{\pm} &=& j(\nu) \pm i \nu, \\
\gamma_0 &=&  \frac{\Gamma(\De_1 - 1)\Gamma(\De_1) \Gamma(\De_2 - 1) \Gamma(\De_2)}{ \gamma(\nu) \gamma(-\nu) }.
\ee
Finally we obtain the anomalous dimensions
\be
\gamma_{h,\bar{h}} &=&  -2\Gamma(\De_1 - 1)\Gamma(\De_1) \Gamma(\De_2 - 1) \Gamma(\De_2) \nn\\
&& \times \Re \int d\nu \frac{\alpha(\nu)}{\gamma(\nu)\gamma(-\nu)}(h\bar{h})^{j(\nu)-1} \left(  \frac{\nu h \bar{h}\sin(\nu \log(h/\bar{h}))} {2\pi^2(h^2-\bar{h}^2)} \right) \nn\\ 
&=& -2\Gamma(\De_1 - 1)\Gamma(\De_1) \Gamma(\De_2 - 1) \Gamma(\De_2) \nn\\
&& \times \Re \int d\nu \frac{\alpha(\nu)}{\gamma(\nu)\gamma(-\nu)}  \left(h\bar{h}\right)^{j(\nu)-1} \Omega_{i\nu}(\log(h/\bar{h})). \label{eq:DeltaG} 
\ee
Up to an overall prefactor, this simply corresponds to taking the amplitude and making the substitutions $\sigma \rightarrow 1/\bar{h}^2$, $\eta \rightarrow \bar{h}^2/h^2$, and $\alpha(\nu) \rightarrow \alpha(\nu)/(\gamma(\nu)\gamma(-\nu))$.

Repeating the same procedure for $\delta P_{h,\bar{h}}$ we find
\be
\delta P_{h,\bar{h}}
&=&  -2\pi \Gamma(\De_1 - 1)\Gamma(\De_1) \Gamma(\De_2 - 1) \Gamma(\De_2) \nn\\
&& \times \Im \int d\nu \frac{\alpha(\nu)}{\gamma(\nu)\gamma(-\nu)}  \left(\frac{1}{h\bar{h}}\right)^{1-j(\nu)} \Omega_{i\nu}(\log(h/\bar{h})). \label{eq:DeltaP}
\ee
Note that the integrals in (\ref{eq:DeltaG}) and (\ref{eq:DeltaP}) do not receive contributions from double-trace states in the $\psi\psi$ channel. The double-trace poles in the $\gamma(\nu)\gamma(-\nu)$ factor in $\alpha(\nu)$, as given in (\ref{eq:alpha}), are explicitly cancelled.

\subsection{Examples}
\subsubsection*{The gravity limit}
We first check whether the general results above reproduce the gravity limit. In 4$d$ we have \cite{Costa:2012cb}: 
\be
\alpha(\nu) &=& \frac{4\pi}{N^2} \frac{1}{4+\nu^2} \frac{\Gamma(\Delta_1 + i\nu/2) \Gamma(\Delta_1 - i\nu/2) \Gamma(\Delta_2 + i\nu/2) \Gamma(\Delta_2 - i\nu/2) }{\Gamma(\Delta_1)\Gamma(\Delta_1-1)\Gamma(\Delta_2)\Gamma(\Delta_2-1)} .
\label{eq:alpha4d}
\ee
Plugging this into (\ref{eq:DeltaG}), we get: 
\be
\gamma_{h,\bar{h}} =  -\frac{8\pi}{N^2} (h\bar{h}) \Re \int d\nu \frac{1}{4+\nu^2} \Omega_{i\nu}(\log(h/\bar{h})) .
\ee
We are left with poles at $\nu = \pm 2i$. The result is simply:
\be
\gamma_{h,\bar{h}} &=&  -\frac{4}{N^2} \frac{ \bar{h}^4}{ h^2 - \bar{h}^2}
\ee
which, after using the correspondence $40N^{2}=C_{T}$, is precisely what we expect for stress tensor exchange.

\subsubsection*{The non-local regime}

We can push our analysis into the non-local regime by further increasing the scattering energy. If the bulk dual is described by a string theory, then for high-energy scattering we expect the string to spread a transverse distance given by
\be
b_I^{2}= \frac{1}{\Delta_{gap}^2} \log s .
\ee
Therefore, non-local effects should appear when the impact parameter $b$ is smaller than $b_I$. In this section, we do not assume the bulk to be a string theory, but we will show that this scale emerges naturally in CFTs with a large gap. 

In the CFT kinematics, the energy squared $s$ and the impact parameter $b$ of the bulk scattering are given by
\be
s= 4h \bar{h}\sim1/\s,\hspace{1cm} b=\log\frac{h}{\bar{h}}\sim-\frac{1}{2}\log\eta .
\ee
In analogy with the bulk impact parameter variables, we introduce the variables $\rho$ and $\rho_I$:
\bea
\rho= -\frac{1}{2}\log\eta , \qquad \rho_{I}^{2}=-\frac{1}{\Delta_{gap}^2} \log \s .
\eea

Generically, the dynamics could be complicated when $\rho \sim \rho_I$. However, there exists a regime where $\rho \ll \rho_I$ such that a saddle point approximation will work to provide simple results. In particular, to evaluate (\ref{eq:4ptRegge}) and (\ref{eq:DeltaG}), we need the factor $\s^{1-j(\nu)}$ to be steep enough such that the integral is dominated by contributions around the maximum of $j(\nu)$. In $\mathcal{N}=4$ SYM, at both weak and strong coupling, the maximum is at $\nu=0$ \cite{Costa:2012cb}. We will assume this is true in the following analysis, but everything can be straightforwardly generalized for a maximum located at a non-zero $\nu$. The validity of this approximation requires $\s^{-1}$ to be large, but not too large to overwhelm the $C_{T}^{-1}$ suppression. In terms of the CFT parameters this is the regime
\be
C_T \gg \frac{1}{\sigma},\hspace{1cm} j^{\prime\prime}(0) \log\sigma \gg 1 .
\label{eq:saddleRegime}
\ee
Applying a saddle point approximation to (\ref{eq:4ptRegge}) then yields
\be
\mathcal{A}(\sigma,\eta) \approx \frac{i \alpha(0)}{\sqrt{2\pi}} \sigma^{1-j(0)}  \frac{(\sqrt{\eta})^{2-j(0)}\log\eta}{(1 - \eta )} \frac{1}{(j''(0) \log\sigma)^{\frac{3}{2}}} \left( 1 + \cO\left(\frac{1}{(-\log\sigma)}\right) \right). \quad \label{eq:saddle}
\ee 
Note that this result is non-singular as $\eta\rightarrow1$, as opposed to the contribution from stress tensor exchange alone. More importantly, due to the appearance of the $\log(\sigma)$ term and the generically fractional value of $j(0)$, we can no longer interpret this correlator as arising from a finite number of single-trace exchanges.

We can make more progress by specializing to CFTs with a large gap in their single-trace spectrum, $\Delta_{gap}\gg 1$. The existence of the stress tensor implies $j(\pm 2i )=2$ and the existence of a large gap implies the leading $\nu^{2}$ term comes with a small coefficient. That is, we should have
\bea
j(\nu)=2-\frac{(4+\nu^{2})}{2\Delta_{gap}^{2}}f(\nu,\Delta_{gap}),
\label{eq:jnuForm}
\eea
where $f(\nu,\Delta_{gap})$ is some unknown function that is regular at $\nu=\pm2i$. Since $j(\nu)$ is finite at $\nu=0$, $f(\nu,\Delta_{gap})$ cannot contain inverse powers of $\nu$. Following \cite{Cornalba:2007fs,Costa:2012cb}, at large $\Delta_{gap}$ we can write down a general form for $f$:
\bea
f(\nu,\Delta_{gap})=\sum\limits_{n=0}^{\infty}\frac{f_{n}(\nu^{2})}{\Delta_{gap}^{2n}}, \hspace{2cm}
f_{n}(\nu^{2})=\sum\limits_{k=0}^{n}c_{n,k}\nu^{2k}.
\eea
Assuming that the CFT is dual to a theory in AdS, consistency with the flat space results of \cite{Caron-Huot:2016icg} says $c_{n,n}=0$ for $n>0$. To see this, we take the flat space limit where $\Delta_{gap}\sim\nu \sim R$, where $R$ is the radius of curvature, and take $R\rightarrow \infty$. Then this is the only term with equal powers of $\nu$ and $\Delta_{gap}$ allowed if we require that the flat space Regge trajectory $j(t)$ is asymptotically linear at large $t$ \cite{Cornalba:2007fs}. These conditions fix $j(\nu)$ to be the answer given in $\mathcal{N}=4$ SYM, where 
\begin{align}
j(\nu)=2-\frac{(4+\nu^{2})}{2\sqrt{\lambda}}+\mathcal{O}(\lambda^{-1})
\end{align}
and $\lambda\sim\Delta_{gap}^{4}$.

It is natural to ask how we can see this from the bootstrap and conformal Regge theory. We start by considering the smallest $m$ such that $c_{m,m}\neq0$. Due to the definition of $\Delta_{gap}$ there must be at least one non-zero $c_{m,m}$ and in $\mathcal{N}=4$ SYM it is $m=0$. Looking at (\ref{eq:alpha}), there is one combination of functions which is sensitive to the decoupling of higher spin states when we take $\Delta_{gap}$ large. For a general $m$ we find:
\bea
\underset{\Delta_{gap}\rightarrow\infty}{\lim} \ \frac{j'(\nu)}{\nu \sin\pi j(\nu)/2)}= -\frac{4}{\pi  \left(\nu ^2+4\right)}-\frac{2 m}{\pi  \nu ^2} . \label{eq:alphalim}
\eea

The other terms in (\ref{eq:alpha}) go to finite numbers when $\Delta_{gap}\rightarrow\infty$, or $j(\nu)\rightarrow 2$. For example, the OPE coefficients $C_{\f\f j(\nu)}|_{j(\nu)=2}$ are fixed by the stress-tensor Ward identity. It is important to note that the $\nu^{-2}$ in (\ref{eq:alphalim}) does not actually lead to new poles in the $\nu$ integral since $\Omega_{i\nu}(\rho)\sim \nu^{2}$ at small $\nu$. However, it will change the values of the residues for the poles generated by the $\gamma(\nu)\gamma(-\nu)$ term in (\ref{eq:4ptRegge}) and therefore change the full correlator. If we require that the pure gravity answer (\ref{eq:alpha4d}) is reproduced in the strong coupling limit, we must have $m=0$. Therefore at large $\Delta_{gap}$ the form of $j(\nu,\Delta_{gap})$ found in $\mathcal{N}=4$ is universal. This condition does not impose that $c_{n,n}=0$ for $n>0$, which would require considering the flat space limit in more detail or the asymptotic Regge limit \cite{Caron-Huot:2016icg}.

The general trajectories with $m\neq0$ however will not affect the calculations for the cross channel data, (\ref{eq:DeltaG}) and (\ref{eq:DeltaP}). There the poles generated by $\gamma(\nu)$ are explicitly cancelled and, as already mentioned, having $m$ non-zero will not introduce any new poles. It is important to note that those results for the crossed channel data were derived assuming we are doing separate integrals over $h$ and $\bar{h}$, or that we are integrating over both $n$ and $j$. That is, the anomalous dimensions for $n,j \gg 1$ will be unchanged, although some low spin anomalous dimensions may be affected.

In the following analysis, we will take $f(0,0)=1$ in (\ref{eq:jnuForm}) to reduce cluttering. It can be easily restored in all our results if needed. Then for CFTs whose central charge is exponentially larger than the gap, $\log C_T\gg\Delta_{gap}^2 \gg 1$, there exist a kinematic regime that satisfies (\ref{eq:saddleRegime}): 
\be
C_T \gg \frac{1}{\sigma},\hspace{1cm} \frac{|\log(\s)|}{\Delta_{gap}^{2}}\gg 1 .
\label{eq:saddleRegime2}
\ee

Note that the scale $\rho_I(\sigma)^{2}=\frac{|\log\sigma|}{\Delta_{gap}^2}$ emerges naturally in the second condition for the saddle point approximation to hold. Taking into account the $\sin(\nu\rho)$ factor in $\Omega_{i\nu}(\rho)$, we find the saddle points are located at:
\bea
\nu_{\pm}=\pm \frac{i \Delta_{gap}^{2}\log(\eta)}{2\log(\s)}=\pm 2i\frac{\rho}{\rho_{I}(\s)^{2}}.
\eea 

The phase of $\alpha(\nu)$ is given by $e^{-i\pi j(\nu)/2}$, so its variation with $\nu$ is suppressed by $\Delta_{gap}^{-2}$ and we can ignore it at this order when finding the saddle.

In order to trust our approximation for $j(\nu)$ we require that $|\nu_{\pm}|\ll\Delta_{gap}$, or
\bea
|\log(\eta)|\ll -\frac{\log(\s)}{\Delta_{gap}} \quad \leftrightarrow \quad \rho\ll \frac{1}{2}\rho_{I}(\s)^{2} \Delta_{gap}.
\eea

We will assume $\nu_{\pm}$ is close enough to the origin so that we can approximate $\alpha(\nu_{\pm})\approx\alpha(0)$. We then find:
\bea
\mathcal{A}(\sigma,\rho)=\frac{i\rho(\sigma e^{-\rho})^{1-j(0)}}{\sqrt{2\pi} \sinh(\rho)}\frac{e^{-\frac{\rho^{2}}{2\rho_{I}(\s)^{2}}}\Delta_{gap}^{3}}{(-\log(\sigma))^{\frac{3}{2}}}\alpha(0).  \label{eq:saddleLargeCoupling}
\label{eq:saddleLargeCoupling}
\eea
 Shifting the location of the saddle gives rise to the extra $e^{-\left(\frac{\rho^{2}}{2\rho_{I}(\sigma)^{2}}\right)}$ factor. This agrees with the behavior expected from a string theory in the bulk \cite{Brower:2007xg,Shenker:2014cwa}. Note that the regime of validity of (\ref{eq:saddle}) does not overlap with that of section \ref{subsec:GravityNLOMatching}, which is $\rho_I\ll\rho$. It also demonstrates that the Regge amplitude is indeed regular when the impact parameter vanishes, $\rho\rightarrow0$.

We can use (\ref{eq:DeltaG}) and (\ref{eq:DeltaP}) to derive the anomalous dimensions and corrections to the OPE coefficients for the double-trace operators in the $\psi\phi$ channel that match to (\ref{eq:saddleLargeCoupling}) under crossing. The result is:  
\be
\gamma_{h,\bar{h}}&=&\Re\frac{\mathit{p}(0)}{\gamma(0)^{2}}\left(2\Delta_{gap}^{3}\right)\frac{\log\frac{h}{\bar{h}}}{\left(2\pi\log h\bar{h}\right)^{\frac{3}{2}}}\frac{\left(h\bar{h}\right)^{2-\frac{2}{\sqrt{\lambda}}}}{h^{2}-\bar{h}^{2}}e^{-\frac{\Delta_{gap}^{2}\left(\log\frac{h}{\bar{h}}\right)^{2}}{2\log h\bar{h}}},\\
\delta P_{h,\bar{h}}&=&\Im\frac{\mathit{p}(0)}{\gamma(0)^{2}}\left(2\Delta_{gap}^{3}\right)\frac{\log\frac{h}{\bar{h}}}{\left(2\pi\log h\bar{h}\right)^{\frac{3}{2}}}\frac{\left(h\bar{h}\right)^{2-\frac{2}{\sqrt{\lambda}}}}{h^{2}-\bar{h}^{2}}e^{-\frac{\Delta_{gap}^{2}\left(\log\frac{h}{\bar{h}}\right)^{2}}{2\log h\bar{h}}},\\
p(\nu)&=&\Gamma(\Delta_{\f}-1)\Gamma(\Delta_{\f})\Gamma(\Delta_{\p}-1)\Gamma(\Delta_{\p})\alpha(\nu).
\ee

In terms of the impact parameter variables $s$ and $b$ we see a similar dependence:
\bea
\{\gamma, \ \delta P\} \propto \frac{ s^{1-2/\sqrt{\lambda}}b}{\sinh(b)\log(s)^{\frac{3}{2}}}e^{-\frac{b^{2}}{2b_{I}(s)^{2}}}\{\cos(\pi j(0)/2),-\sin(\pi j(0)/2)\}, \label{eq:schannelRegge}
\eea 
and the saddle point approximation is valid when $b_{I}\gg 1,b$. Looking at (\ref{eq:schannelRegge}), we see that taking $b\rightarrow 0$ does not lead to any new divergences. This is contrast to when the stress-tensor is dominant, where there are additional singularities when $b\rightarrow 0$, or $\bar{h}\rightarrow h$, see (\ref{eq:adgravity}), (\ref{eq:adSTT}), and (\ref{eq:adMixed}).

\subsection{Chaos Bounds and Eikonalization}

To make a more direct connection to the chaos bound~\cite{Maldacena:2015waa}, it is convenient to make the following change of variables
\bea
\eta=e^{-4\pi x}, \qquad \s = -4 i e^{2\pi(x-t)}.
\eea
The region of parameter space relevant for the chaos bound is $t\gg1$ and $x$ fixed, with a crossover region around the scrambling time $t_{*}=\frac{1}{2\pi}\log(N^{2})$. The analog of their function $f(x,t)$ in our case (with an abuse of notation) is:
\bea
f(x,t)=1+\mathcal{A}(x,t).
\eea
This function obeys all the assumptions of \cite{Maldacena:2015waa}. In particular, it is real when $t$ is real and for $t>0$ we have:
\bea
\frac{1}{1-f}\bigg|\frac{df}{dt}\bigg|\leq 2\pi+\mathcal{O}(e^{-4\pi t}), \hspace{1cm} |f(x,t)|\leq 1 . \label{eq:chaosbound}
\eea
Applying this to (\ref{eq:saddle}) with $t$ real, we obtain the following conditions: 
\be
j(0)\le 2+\frac{3}{4\pi t},\hspace{1cm} \arg(\alpha(0))=\frac{-\pi j(0)}{2}+\pi .
\ee
We see the bound on the Regge intercept is modified because the $\log(\s)^{-\frac{3}{2}}$ in (\ref{eq:saddle}) slightly softens the divergence when we take $\s\rightarrow 0$, or $t$ large. The modification is of order $\log^{-1}(N^{2})$ given the form of the scrambling time $t_{*}$.

Recalling the definition of $\alpha(\nu)$ in (\ref{eq:alpha}) we note it already carries an explicit phase dependence $e^{\frac{-i\pi j(0)}{2}}$. Moreover, the assumption that there is a saddle at $\nu=0$ implies $j''(0)<0$. To complete the phase dependence matching we must fix the sign of $\mathcal{A}$, which will ensure chaos decreases the value of an out-of-time correlator. This gives the final constraint:
\bea
\sin\left(\frac{\pi j(0)}{2}\right)C_{11j(0)}C_{22j(0)}\geq 0.
\eea

These bounds, which constrain how two scalars can couple to the leading Regge trajectory, imply the following sign constraints:
\begin{align}
\textrm{sgn} \ \gamma_{h,\bar{h}}=\cos(\pi j(0)/2), \hspace{1cm}
\textrm{sgn} \ \delta P_{h,\bar{h}}=-\sin(\pi j(0)/2).
\end{align}

For theories like $\mathcal{N}=4$ SYM the dominant saddle is at $\nu=0$ and we have $1\leq j(0,\lambda) \leq 2$, or $\delta P_{h,\bar{h}} \leq 0$ and $\gamma_{h,\bar{h}}\leq 0$.

We can ask how these constraints should be interpreted if we assume the four-point functions eikonalize in the limit $\s\rightarrow 0$ with $\s N^{2}$ fixed. The large $h$ and $\bar{h}$ anomalous dimensions have a well known connection to eikonalization for graviton exchange when $\Delta_{gap}=\infty$. Namely they correspond to the real part of the phase shift and requiring that they are negative is equivalent to requiring AdS causality \cite{Cornalba:2007zb}. Eikonalization has also been argued to occur in AdS for finite $\lambda$, or finite $\Delta_{gap}$, using Pomeron techniques \cite{Brower:2006ea,Brower:2007qh,Brower:2007xg}, in which case the phase shift $e^{-2\pi i \Gamma}$ will have both real and imaginary contributions. The statement that $\delta P_{h,\bar{h}}<0$ then turns into $\Im(\Gamma)<0$, or that we have AdS unitarity \cite{Cornalba:2008qf}. 

From the CFT perspective, proving eikonalization at finite $\Delta_{gap}$ remains an open question. In particular there is the question: in what situations do the corrections to the OPE coefficients eikonalize in a manner similar to the anomalous dimensions? Based off known results in flat space and weakly curved AdS, we can expect that eikonalizing the tree level results gives a good approximation for either very large impact scattering $b\gg b_{I}$, when graviton exchange is dominant, or when $b<b_{I}$ in which case long string creation gives the largest contribution. In string theory there is also an intermediate regime, $b_{I}<b<b_{D}$ where diffractive scattering, or tidal excitations, gives the leading contribution to the imaginary part of the phase shift and the phase shift becomes an operator mapping initial states to final states \cite{Amati:1987wq,Amati:1987uf}. When $b< b_{I}$ this contribution is expected to be suppressed by $\log(N^{2})^{-1}$, and when $b\gg b_{I}$ they will be suppressed by $1/\sqrt{\lambda}$ in comparison to the elastic amplitude~\cite{Shenker:2014cwa}. Therefore, they will give subleading effects for the regions of parameter space we have considered. A fuller understanding of tidal excitations will require going beyond tree level in the bulk dual.

\subsection{New States}
At this point, we can note that there is some tension between the result for $\delta P_{h,\bar{h}}$ in (\ref{eq:DeltaP}) and what we expect from both the Euclidean t-channel OPE and the derivative relation between OPE coefficients and anomalous dimensions. The derivative relation~\cite{Heemskerk:2009pn}, which has been proven for contact diagrams and holds approximately at large $h$ and $\bar{h}$ for exchange Witten diagrams, states:
\bea
2P_{MFT}\delta P_{h,\bar{h}}=(\partial_{h}+\partial_{\bar{h}})P^{MFT}_{h,\bar{h}}\gamma_{h,\bar{h}}.
\label{eq:DerivativeRelation}
\eea
This result implies $\delta P_{h,\bar{h}}$ must grow slower in comparison to $\gamma_{h,\bar{h}}$ by a factor of $\sqrt{h\bar{h}}$. We recall that this asymptotic relation was derived for exchange diagrams using the condition that we do not generate non-OPE singularities on the Euclidean sheet. The crossing equation on the first sheet is:
\begin{align}
\s^{-2\Delta_{\p}}\eta^{-\Delta_{\p}} G(\s,\eta)=&\sum_{h,\bar{h}}P^{MFT}_{h,\bar{h}}\big[ \gamma_{h,\bar{h}}\frac{1}{2}(\partial_{h}+\partial_{\bar{h}})+ \delta P_{h,\bar{h}}\big]g^{a,a}_{h,\bar{h}}(1-\s,1-\eta \s) \nonumber \\ +&\sum_{\mathcal{O'}}P_{\mathcal{O'}}g^{a,a}_{\mathcal{O}'}(1-\s,1-\eta \s).
\end{align}
Consider the contribution from the anomalous dimensions alone. For $\gamma_{h,\bar{h}}\sim (h\bar{h})^{j-1}$, after approximating the sum as an integral, we produce a singularity on the first sheet that grows like $\s^{-2\Delta_{\f}+\frac{3}{2}-j}$ in the limit $\s\rightarrow 0$. This is generically not consistent with the $\psi\psi$-channel OPE, and when $j\geq \frac{3}{2}$ we generate a singularity that cannot be reproduced by the $t$-channel OPE. If we consider stress tensor/graviton exchange in a theory with $\Delta_{gap}=\infty$, then $j=2$ and the second sum over single-trace operators $\mathcal{O}'$ disappears. Then it is clear that to cancel this unphysical singularity, $\delta P$ must grow slower than $h\bar{h}$. In particular, if (\ref{eq:DerivativeRelation}) holds then these two contributions combine to become a total derivative that vanishes after the integration over $h$ and $\bar{h}$. In this limit, (\ref{eq:DeltaP}) gives $\delta P_{h,\bar{h}}=0$, or to be more precise, $\delta P_{h,\bar{H}}$ must grow slower than $h\bar{h}$, which is consistent with the derivative relation.

At large but finite $\Delta_{gap}$, (\ref{eq:DeltaP}) gives $\delta P_{h,\bar{h}}\sim (h\bar{h})^{j(0)-1}$, so the corrected OPE coefficients produce a singularity of the form $\s^{1-j(0)}$. We cannot cancel this singularity in the same way as before since both $\gamma_{h,\bar{h}}$ and $\delta P$ are now fixed to grow at the same rate. Instead, this divergence must be cancelled by the sum over single-trace operators $\mathcal{O}'$ in the $\psi\phi$ channel, or operators that first appear at order $C_{T}^{-1}$.  Requiring that this divergence cancels yields
\bea
-\sum_{h,\bar{h}}P^{MFT}_{h,\bar{h}}\delta P_{h,\bar{h}}g_{h,\bar{h}}(z,\bar{z}) \approx \sum_{\mathcal{O'}}P_{\mathcal{O'}}g_{\mathcal{O}'} (z,\bar{z}),\label{eq:canceldiv}
\eea
where the $``\approx"$ is because we only require that the most singular terms in the limit $\s\rightarrow 0$ match. 

Furthermore, since the theory is unitary we have $P_{\mathcal{O}'} \geq0$. Reflection positivity also guarantees that the s-channel blocks $g_{h,\bar{h}}$ and $g_{\mathcal{O}'}$ are positive, so crossing symmetry on the first sheet gives another way to see why $\delta P_{h,\bar{h}}<0$. Unitarity here only requires $P^{MFT}_{h,\bar{h}}(1+\delta P_{h,\bar{h}})\geq0$, and, since $\delta P$ is $1/N$ suppressed, it can in principle be positive or negative. This formula also aligns with our expectations from AdS unitarity: when the phase shift has an imaginary part at tree level, i.e. $1\leq j(0)<2$, the scattering is no longer purely elastic and we can have absorption. The RHS of (\ref{eq:canceldiv}) then plays the role of the total cross section $\s_{tot}$ in the optical theorem -- it is a manifestly positive quantity which gives a sign constraint on the imaginary part of the phase shift \cite{Brower:2007xg}, or here $\delta P_{h,\bar{h}}$.

Finally, we note that since each s-channel block does not have the requisite power law singularity when $\s\rightarrow 0$, we need to assume that there are an infinite number of new operators $\mathcal{O}'$. It is also crucial to note that generically the new single-trace states $\mathcal{O}'$ do not add in phase, so their contribution is subleading on the second sheet. We currently cannot constrain the spectrum and OPE coefficients of the new operators beyond what their sum should be to cancel an unwanted singularity, although with additional assumptions it might be possible to do so. 

It may not be surprising that when the gap becomes finite in the $\psi\psi$-channel, an infinite family of new single-trace operators also appears in the $\psi\phi$-channel. What crossing symmetry tells us is that we cannot make $\Delta_{gap}$ finite in one channel but effectively infinite in the crossed channel. At large but finite `t Hooft coupling these new states correspond to long string states created in the $\psi\phi$-channel. This aligns with expectations from flat space, high-energy scattering: that when the impact parameter variable $b<b_{I}$ we can have the production of states in the s-channel if the theory contains extended objects \cite{Amati:1987uf,Amati:1987wq,Camanho:2014apa}.

\section{Discussion}
In this work we have studied the analytic bootstrap in the Regge limit, which is related to high-energy, fixed impact parameter scattering in the AdS dual. We have re-derived results for anomalous dimensions in theories with a parametrically large gap, obtained new results for double-trace operators when the spin of the single-trace operators is unbounded, and derived new constraints on both analytically continued t-channel OPE coefficients and s-channel double-trace data using the chaos bound. These constraints imply that in the bulk dual the exchange of the leading Regge trajectory leads to a universally attractive force between two scalar particles and that the theory in AdS obeys bulk causality and unitarity.

In this work we have focused on correlation functions of four scalar operators and two currents and two scalars $\langle J \phi\phi J\rangle$. An important direction forward is a more thorough analysis for external operators with spin. In particular, by studying $\<TTTT\>$ in theories with a parametrically large gap it should be possible to derive the $a=c$ constraint and see how this bound is corrected as we start to include $\Delta_{gap}^{-1}$ corrections. It would be interesting to make a connection to \cite{Afkhami-Jeddi:2016ntf}. In this work we were able to project out the contributions from the t-channel double-trace contributions when calculating the s-channel anomalous dimensions. It is also interesting to consider if there are alternative methods to project out these operators at the level of the correlator.

Such an analysis is crucial in order to expand our understanding of the universality of Einstein gravity, as well as finding the properties of the leading Regge trajectory are truly universal. In this work we have also derived constraints on $\alpha(\nu)$ and $j(\nu)$ around $\nu=0$ for the leading Regge trajectory, which match expectations from AdS/CFT \cite{Brower:2006ea,Brower:2007xg}. It also interesting to ask how much more can be derived about the spectrum of the leading Regge trajectory. Can we derive constraints on their asymptotic behavior for large $\nu$? In $\mathcal{N}=4$ SYM there is a qualitative change for the operators on the leading Regge trajectory $\mathcal{O}_{\Delta,j}$ from $j\ll \sqrt{\lambda}$, where $\Delta\sim\lambda^{\frac{1}{4}}\sqrt{j}$, to $j\gg\sqrt{\lambda}$ where $\Delta-j\sim\sqrt{\lambda}\log(j/\sqrt{\lambda})$ \cite{Gubser:2002tv}. It is an open question if these results can also be derived using bootstrap techniques.

The bootstrap in the Regge limit also has a close connection to known results for high-energy, fixed impact parameter scattering, both when the dual theory is pure gravity and when it is a weakly coupled string theory. For both cases, we have provided additional evidence that the anomalous dimensions of double-trace operators map onto the real part of the phase shift. Furthermore, the general structure we observe, both here and in the lightcone limit \cite{Li:2015itl,Hofman:2016awc}, is that decomposing a correlation function in terms of the Lorentz representations of the double-trace operators corresponds to diagonalizing the phase shift matrix. When $j(0)$ is no longer exactly 2 we see that crossing symmetry on the first sheet implies the existence of an infinite number of new single-trace operators. This matches our expectations from high-energy string scattering in flat space where in the corresponding regime we have the production of string states in the s-channel \cite{Amati:1987wq,Amati:1987uf}. In order to complete this dictionary we need to understand how to see tidal excitations of the string from the bootstrap. This will require going beyond tree level and understanding how to derive eikonalization from the CFT away from the pure gravity limit \cite{Brower:2007xg}.

A surprising new result to come out of the bootstrap is a direct connection between the results of the lightcone bootstrap and the spectrum of low central charge CFTs like the Ising model \cite{Alday:2015ewa, Alday:2015ota, Simmons-Duffin:2016wlq}. The success of this work can be explained by the recent proof of a CFT version of the Froissart-Gribov formula which explains why operators with spin $j\geq 2$ are organized in analytic families \cite{Caron-Huot:2017vep}. This proof explicitly relied on the correlation function having nice behavior in the Regge limit. It is of clear future interest and importance to understand the interplay of analytic and numerical techniques and if the study of correlation functions in the Regge limit can shed new light on the numerical bootstrap.

\section*{Acknowledgements}
We thank Simon Caron-Huot, Tom Faulkner, Liam Fitzpatrick, Tom Hartman, Diego Hofman, Jared Kaplan, Sandipan Kundu, Jo\~ao Penedones, Eric Perlmutter, Fernando Rejon-Barrera, David Simmons-Duffin, Matthew Walters, Junpu Wang, and Sasha Zhiboedov for discussions. DP and DM are supported by NSF grant PHY-1350180 and Simons Foundation grant 488651. DM thanks Princeton University for its hospitality during the completion of this work.

\appendix
\section{Conventions and Definitions}
\label{App:Conventions}
The conformal blocks are given by: 
\begin{align}
&g^{(d=2),a,b}_{h,\bar{h}}(z,\bar{z})=k_{h}(z)k_{\bar{h}}(\bar{z})+(z\leftrightarrow \bar{z}), \\
&g^{(d=4),a,b}_{h,\bar{h}}(z,\bar{z})=\frac{z\bar{z}}{z-\bar{z}}k_{h}(z)k_{\bar{h}-1}(\bar{z})+(z\leftrightarrow \bar{z}), \\
&k_{h}(z)=z^{h} \ _{2}F_{1}(h+a,h+b,2h,z). 
\end{align}
Some useful formulas for conformal Regge theory not given in the body of the paper are \cite{Costa:2012cb}:
\begin{align}
&K_{\De,J} = \frac{\Gamma(\De+J) \Gamma(\De-h+1) (\De-1)_J}{4^{J-1} \Gamma(\frac{\De+J}{2})^4 \Gamma(\frac{2\De_1-\De+J}{2})\Gamma(\frac{2\De_2-\De+J}{2})\Gamma(\frac{2\De_1+\De+J-d}{2})\Gamma(\frac{2\De_2+\De+J-d}{2})}, \\
&\Omega_{i\nu}(\rho) = \frac{\nu \sinh(\pi\nu) \Gamma(h-1+i\nu) \Gamma(h-1-i\nu) {}_2F_1(h-1+i\nu,h-1-i\nu,h-1/2;-\sinh^2(\rho/2))}{2^{2h-1} \pi^{h+1/2} \Gamma(h-1/2)}.
\end{align}

The approximations of the hypergeometrics needed for the Regge limit will be the same those used in the lightcone limit,
and we find that the 2d and 4d t-channel blocks can be approximated by Bessel functions. For example, in $d=2$ the block is approximated as
\begin{align}
g^{(d=2),a,b}_{h,\bar{h}}(z,\bar{z})\approx \frac{\sqrt{h\bar{h}}}{\pi}2^{2(h+\bar{h})}K_{a+b}(2h\sqrt{1-z})K_{a+b}(2\bar{h}\sqrt{1-\bar{z}})((1-z)(1-\bar{z}))^{\frac{1}{2}(a+b)}+(z\leftrightarrow \bar{z}),  \label{eq:2dBesselApp} 
\end{align}
while in $d=4$ we have
\begin{align}
g^{(d=4),a,b}_{h,\bar{h}}(z,\bar{z})\approx \frac{\sqrt{h\bar{h}}}{\pi}2^{2(h+\bar{h}-1)}\frac{1}{z-\bar{z}}K_{a+b}(2h\sqrt{1-z})K_{a+b}(2\bar{h}\sqrt{1-\bar{z}})((1-z)(1-\bar{z}))^{\frac{1}{2}(a+b)}+(z\leftrightarrow \bar{z}),  \label{eq:4dBesselApp}  
\end{align}
with $a=-\frac{1}{2}(\Delta_{1}-\Delta_{2})$ and $b=\frac{1}{2}(\Delta_{3}-\Delta_{4})$.

It is possible to derive the approximate form of the crossed channel blocks in any even dimension, since we know them in closed form and can apply the usual Bessel function approximations to the hypergeometrics. We do not have similar closed form expressions in odd dimensions, although in \cite{Cornalba:2006xm,Cornalba:2007zb} they presented a simple formula for these blocks in all dimensions using an impact parameter formalism that makes the connection to high-energy AdS scattering manifest. The impact parameter blocks in $d=2$ agree with (\ref{eq:2dBesselApp}), although in general it is only known that these impact parameter blocks satisfy the correct quadratic Casimir equation. In appendix \ref{App:Impact} we will show that when matching the $1/N^{2}$ corrections, the conformal and impact parameter blocks will always agree in $d=4$ and will agree when matching operators of integer twist in any dimension.

The t-channel blocks in the Regge limit for general dimension and $d=4$ are given by
\bea
g^{Regge}_{\Delta,j}(\s,\eta)=2\pi i \s^{1-j}\eta^{\frac{1}{2}(\Delta-j)}\frac{\Gamma (\Delta +j-1) \Gamma (\Delta +j)}{\Gamma \left(\frac{\Delta +j}{2}\right)^4} {}_{2}F_{1}\bigg(\frac{d-2}{2},\Delta-1,\Delta-\frac{d-2}{2},\eta\bigg),
\eea
and
\bea
g^{Regge,d=4}_{\Delta,2}(\s,\eta)=\frac{2 i \pi  \sigma ^{1-j} \eta ^{\frac{\Delta -j}{2}} \Gamma (\Delta +j-1) \Gamma (\Delta +j)}{(1-\eta ) \Gamma \left(\frac{\Delta +j}{2}\right)^4}. \label{eq:ScalarSpinjApp}
\eea

\section{Integrals of Bessel Functions}
\label{App:IntK}

The general integrals needed when solving the bootstrap equations are:
\begin{align}
&I_{1}(a_{1},a_{2},b,z,\bar{z})=\int_{0}^{\infty} d\bar{h}\int_{\bar{h}}^{\infty}dh h^{a_{1}}\bar{h}^{a_{2}}K_{b}(2h\sqrt{\bar{z}})K_{b}(2\bar{h}\sqrt{z}) \nonumber
\\ \nonumber
&=\frac{1}{16} \bar{z}^{-\frac{a_{1}}{2}-\frac{1}{2}} z^{-\frac{a_{2}}{2}-\frac{1}{2}}\bigg(\Gamma (\frac{1}{2} (a_{1}-b+1)) \Gamma (\frac{1}{2} (a_{1}+b+1)) \Gamma (\frac{1}{2} (a_{2}-b+1)) \Gamma (\frac{1}{2} (a_{2}+b+1))- 
\\ \nonumber & \frac{1}{a_{1}-b+1}2 \Gamma (b) \Gamma (\frac{1}{2} (a_{1}+a_{2}+2)) \gamma ^{a_{1}-b+1} \Gamma (\frac{1}{2} (a_{1}+a_{2}-2 b+2)) \, \times 
\\ \nonumber&  _3F_2(\frac{a_{1}}{2}+\frac{a_{2}}{2}+1,\frac{a_{1}}{2}+\frac{a_{2}}{2}-b+1,\frac{a_{1}}{2}-\frac{b}{2}+\frac{1}{2};1-b,\frac{a_{1}}{2}-\frac{b}{2}+\frac{3}{2};\gamma ^2)-
\\ \nonumber & \frac{1}{a_{1}+b+1}2 \Gamma (-b) \Gamma (\frac{1}{2} (a_{1}+a_{2}+2)) \gamma ^{a_{1}+b+1} \Gamma (\frac{1}{2} (a_{1}+a_{2}+2 b+2)) \, \times
\\ & _3F_2(\frac{a_{1}}{2}+\frac{a_{2}}{2}+1,\frac{a_{1}}{2}+\frac{b}{2}+\frac{1}{2},\frac{a_{1}}{2}+\frac{a_{2}}{2}+b+1;\frac{a_{1}}{2}+\frac{b}{2}+\frac{3}{2},b+1;\gamma ^2)\bigg),
\end{align}

\begin{align}
& \nonumber I_{2}(a_{1},a_{2},b,z,\bar{z})=\int_{0}^{\infty} d\bar{h}\int_{\bar{h}}^{\infty}dh h^{a_{1}}\bar{h}^{a_{2}}K_{b}(2h\sqrt{z})K_{b}(2\bar{h}\sqrt{\bar{z}})
\\ \nonumber &\frac{1}{8} z^{-\frac{1}{2} (a_{1}+1)} \bar{z}^{-\frac{1}{2} (a_{2}+1)} \Gamma \left(\frac{1}{2} (a_{1}+a_{2}+2)\right) 
\\ \nonumber &\bigg(\frac{\gamma ^{a_{2}-b+1} \Gamma (b) \Gamma \left(\frac{1}{2} (a_{1}+a_{2}-2 b+2)\right) \, _3F_2\left(\frac{a_{1}}{2}+\frac{a_{2}}{2}+1,\frac{a_{1}}{2}+\frac{a_{2}}{2}-b+1,\frac{a_{2}}{2}-\frac{b}{2}+\frac{1}{2};1-b,\frac{a_{2}}{2}-\frac{b}{2}+\frac{3}{2};\gamma ^2\right)}{a_{2}-b+1}
\\ & \hspace{-.5cm}+\frac{\gamma ^{a_{2}+b+1} \Gamma (-b) \Gamma \left(\frac{1}{2} (a_{1}+a_{2}+2 b+2)\right) \, _3F_2\left(\frac{a_{1}}{2}+\frac{a_{2}}{2}+1,\frac{a_{2}}{2}+\frac{b}{2}+\frac{1}{2},\frac{a_{1}}{2}+\frac{a_{2}}{2}+b+1;\frac{a_{2}}{2}+\frac{b}{2}+\frac{3}{2},b+1;\gamma ^2\right)}{a_{2}+b+1}\bigg),
\end{align}
where $\gamma=\sqrt{\frac{\bar{z}}{z}}$ and $\gamma<1$. There is an asymmetry between the two integrals since we have $0<\bar{z}<z<1$ and $\bar{h}\leq h$. The $_{3}F_{2}$ terms in the above integrals will match double-trace operators that appear in the direct channel decomposition of exchange Witten diagrams.

The simpler integral we need in matching just the single-trace exchange term is given by:
\begin{align}
I_{3}(a,b,z)=\int_{0}^{\infty}dh h^{a}K_{b}(2h\sqrt{z})=\frac{1}{4} z^{-\frac{a}{2}-\frac{1}{2}} \Gamma \left(\frac{1}{2} (a-b+1)\right) \Gamma \left(\frac{1}{2} (a+b+1)\right).
\end{align}

\section{Impact Parameter Formalism}
\label{App:Impact}
\subsection{Definitions}
In this appendix we will review the impact parameter formalism of \cite{Cornalba:2006xm,Cornalba:2007zb,Cornalba:2007fs} and its connection to the standard conformal blocks for a generic correlation function of distinct scalars. Our results for the impact parameter blocks will differ slightly because we consider the ordering $\<\f_{1}\f_{1}\f_{2}\f_{2}\>$ instead of $\<\f_{1}\f_{2}\f_{1}\f_{2}\>$ and work with different conventions for the cross ratios. We start by defining some conventions. We will work in Minkowski space $\mathds{M}$ with a mostly minus metric. The future Milne wedge $M$ is given by $x^{2}<0$ and $x^{0}>0$. The hyperbolic subspace $H_{d-1}$ of $M$ is given by $x^{2}=-1$. The past Milne wedge and corresponding hyperbolic subspace are denoted by $-M$ and $-H_{d-1}$. Finally we will parametrize the cross ratios as:
\bea
z\bar{z}=q^{2}p^{2}, \qquad z+\bar{z}=2p\cdot q ,
\eea
where $p$ and $q$ are points in $-M$. In \cite{Cornalba:2006xm}, in analogy with flat space partial waves, they introduced the impact parameter blocks for $\<\f_{1}\f_{1}\f_{2}\f_{2}\>$:
\begin{align}
\mathcal{I}^{Identity}_{h,\bar{h}} =& \,\mathcal{N}_{\Delta_{1}}\mathcal{N}_{\Delta_{2}}(-q^{2})^{\Delta_{1}-\Delta_{2}} \nn\\
&\times \int_{M}\frac{dx}{|x|^{d-2\Delta_{1}}}4h\bar{h}e^{-2q\cdot x}\int_{M}dy \frac{1}{|y|^{d-2\Delta_{2}}} e^{-2p\cdot y}\delta(2y\cdot x+h^{2}+\bar{h}^{2})\delta (x^{2}y^{2}-h^{2}\bar{h}^{2}) \label{eq:genImpDr},\nn\\\\
\mathcal{N}_{\Delta}=&\,\frac{2\pi^{1-\frac{d}{2}}}{\Gamma(\Delta)\Gamma(1+\Delta-\frac{d}{2})}.
\end{align}

Using the integral
\bea
\mathcal{N}_{\Delta}\int_{M}\frac{dx}{|x|^{d-2\Delta}}e^{-2p\cdot x}=\frac{1}{|p|^{2\Delta}} ,\label{eq:identity}
\eea
where $\int_{M}dy=\int_{0}^{\infty}r^{d-1}dr\int_{H_{d-1}}\widetilde{dy}$, one can show 
\bea
\int_{0}^{\infty} dh\int_{0}^{h} d\bar{h} \ \mathcal{I}^{Identity}_{h,\bar{h}}=(z\bar{z})^{-\Delta_{2}} 
\eea
as expected for identity matching.

It was shown in \cite{Cornalba:2006xm} that this function satisfies the quadratic Casimir differential equation in the s-channel Regge limit with $h\sim\bar{h}\sim z^{-1/2}\sim\bar{z}^{-1/2}$:
\bea
D^{t}\mathcal{I}^{Identity}_{h,\bar{h}}=(h^{2}+\bar{h}^{2})\mathcal{I}^{Identity}_{h,\bar{h}},
\\
D^{t}=z\partial_{z}^{2}+\bar{z}\partial_{\bar{z}}^{2}+(a+b+1)(\partial_{z}+\partial_{\bar{z}})+\frac{d-2}{z-\bar{z}}(z\partial_{z}-\bar{z}\partial_{\bar{z}}) ,\label{eq:tRegge}
\eea
with $a=-\frac{1}{2}(\Delta_{1}-\Delta_{2})$ and $b=\frac{1}{2}(\Delta_{3}-\Delta_{4})$. 

However, (\ref{eq:identity}) is not quite what we want since it corresponds to blocks for a specific kind of correlator, i.e. when there are two pairs of identical scalars so there is an identity contribution in one channel. Moreover, it corresponds to conformal blocks dressed by the MFT OPE coefficients. To fix this we can divide (\ref{eq:identity}) by the OPE coefficients in (\ref{eqn:MFT}) and note that the resulting equation is a function of $\Delta_{2}-\Delta_{1}=a+b$ alone. The t-channel differential operator $D^{t}$ is also only a function of $a+b$ alone, so we can simply make the replacement $\Delta_{2}-\Delta_{1}\rightarrow a+b$ everywhere to find:
\begin{align}
\mathcal{I}^{t,(a,b)}_{h,\bar{h}}=&2^{2(h+\bar{h})}2^{-d} \pi ^{1-d}(-q^{2})^{-a-b}\nonumber \\& \times\int_{M}\frac{dx}{|x|^{\frac{1}{2}+a+b}}\frac{dy}{|y|^{\frac{1}{2}-a-b}}e^{-2q\cdot x-2p\cdot y}\delta(2y\cdot x+h^{2}+\bar{h}^{2})\delta(x^{2}y^{2}-h^{2}\bar{h}^{2})\frac{4h\bar{h}}{(h^{2}-\bar{h}^{2})^{\frac{d}{2}-1}}. \qquad  \nonumber \\
\label{eq:genImp}
\end{align}

We have used the delta functions to convert between the ${x,y}$ basis and the $h$, $\bar{h}$ basis, choosing to leave some factors of the latter explicit for simplicity later. In practice we will always look at the case $a=b$, but it is necessary to have the general formula when constructing spinning conformal blocks.

\subsection{d=4}
We will now show that when integrating the impact parameter and conformal blocks against the function
\bea
\int d\nu 2^{-2(h+\bar{h})}(h\bar{h})^{c}(h^{2}-\bar{h}^{2})^{d/2-1}\Omega_{i\nu}\left(\log\bigg(\frac{h}{\bar{h}}\bigg)\right)\beta(\nu)\label{eq:Ansatz4d},
\eea
we will get the same answer in both cases. The above ansatz is the most general one for the MFT OPE coefficients multiplied by the anomalous dimensions.

Before we start, we list some useful formulas:
\begin{align}
&\frac{1}{(-2x\cdot y)^{\eta}}=\int d\nu V(\nu,\eta)\Omega_{i\nu}(x,y), \quad V(\nu,\eta)=\frac{\pi^{\frac{d}{2}-1}}{2}\frac{\Gamma(\frac{\eta-d/2+1+i\nu}{2})\Gamma(\frac{\eta-d/2+1-i\nu}{2})}{\Gamma(\eta)}, \quad x,y\in H_{d-1}  \label{eq:powerharmonic}
\\
&\int_{H_{d-1}}dw \ \Omega_{i\nu}(w,w')\Omega_{i\bar{\nu}}(w',w'')=\frac{1}{2}(\delta(\nu-\bar{\nu})+\delta(\nu+\bar{\nu}))\Omega_{i\nu}(w,w'') .\label{eq:orthogonal}
\end{align}
The integral representation of a power law (\ref{eq:powerharmonic}) is one that will show up repeatedly later, while (\ref{eq:orthogonal}) states the harmonic functions $\Omega_{i\nu}(w,w')$ are a complete basis of functions on $H_{d-1}$. To be precise, $\Omega_{i\nu}(w,w')$ depends on the geodesic distance between $w/|w|$ and $w'/|w'|$ on $H_{d-1}$.

We start by integrating (\ref{eq:Ansatz4d}) against the general impact parameter blocks 
\bea
I_{impact}=2^{-d} \pi ^{1-d}|q|^{-2(a+b)}\int_{M} \frac{dx}{|x|^{1/2+a+b-c}}\frac{dy}{|y|^{1/2-a-b-c}}\int d\nu \Omega_{i\nu}(x,y)\beta(\nu)e^{-2q\cdot x -2p\cdot y}.
\eea

We can plug in $x=|x|\tilde{x}$, $y=|y|\tilde{y}$, $p=-|p|\tilde{p}$, and $q=-|q|\tilde{q}$, where all the vectors with tildes are in $H_{d-1}$ and do the radial integrals to obtain:
\bea
I_{impact}=2^{-d} \pi ^{1-d}\int d\nu \int_{H_{d-1}} \tilde{dx}\tilde{dy}|p|^{-a-b-c-d+\frac{1}{2}} |q|^{-a-b-c-d+\frac{1}{2}} \Gamma \left(-a-b+c+d-\frac{1}{2}\right) \nonumber \\ \Gamma \left(a+b+c+d-\frac{1}{2}\right) 
 \frac{\Omega_{i\nu}(\tilde{x},\tilde{y})}{(-2\tilde{p}\cdot\tilde{y})^{a+b+c+d-\frac{1}{2}} (-2\tilde{q}\cdot\tilde{x})^{-a-b+c+d-\frac{1}{2}}}\beta(\nu)e^{-2q\cdot x -2p\cdot y}. \quad
\eea
Finally we will use (\ref{eq:powerharmonic}) twice and (\ref{eq:orthogonal}) to do the integrals over hyperbolic space and use the resulting delta functions to do the $\nu$ integrals. The answer is, after reverting to the cross ratios $z$ and $\eta=\bar{z}/z$,
\bea
I_{impact}&=&2^{-d} \pi ^{1-d}\eta^{\frac{1}{2}(-a-b-c-d+\frac{1}{2})}z^{-a-b-c-d+\frac{1}{2}}\int d\nu V(\nu,e_{1})V(\nu,e_{2})\Gamma(e_{1})\Gamma(e_{2})\Omega_{i\nu}\left(-\frac{1}{2} \log\big(\eta\big)\right)\beta(\nu). \nn\\ \label{eq:impactfinal}
\eea

To compare with the 4d blocks we can either close the contour in $\nu$ and sum over all the poles or do the comparison under the $\nu$ integral. We will take the latter approach. We now need to evaluate the following integral:
\begin{align}
I_{4d}&=\int d\nu dh d\bar{h} \ 2^{-2(h+\bar{h})}(h\bar{h})^{c}(h^{2}-\bar{h}^{2})\Omega_{i\nu}\big(\log(h/\bar{h})\big)\beta(\nu) g^{4d}_{h,\bar{h}}(z,\bar{z}) \nonumber
\\
&=\int d\nu dh d\bar{h} \ i\nu(h\bar{h})^{c+3/2}\frac{1}{16\pi^{3}}\frac{1}{z-\bar{z}}K_{a+b}(2h\sqrt{\bar{z}})K_{a+b}(2\bar{h}\sqrt{h})(z\bar{z})^{-\frac{1}{2}(a+b)}\beta(\nu)\bigg[\left(\frac{h}{\bar{h}}\right)^{-i \nu }-\left(\frac{h}{\bar{h}}\right)^{i \nu }\bigg],
\end{align}
where we used the symmetry of $g^{4d}_{h,\bar{h}}(z,\bar{z})$ in $z$ and $\bar{z}$ so we only need to write down one product of Bessel functions.

We then find:
\begin{align}
I_{4d}&=\int d\nu\frac{-1}{256 \pi ^3 (z-\bar{z})}i \nu  \left(z^{i \nu }-\bar{z}^{i \nu }\right) \Gamma (\frac{1}{2} (- a- b+ c- i \nu +5/2)) \Gamma (\frac{1}{2} ( a+ b+ c- i \nu +5/2)) \nonumber \\ 
&\times \Gamma (\frac{1}{2} (- a- b+ c+ i \nu +5/2)) \Gamma (\frac{1}{2} ( a+ b+ c+i \nu +5/2)) (z \bar{z})^{\frac{1}{2} (- a- b- c- i \nu -5/2)} \beta(\nu) \nonumber
\\
&=\int d\nu \frac{1}{64 \pi }\eta^{\frac{1}{2} (- a- b- c-7/2)}z^{-a-b-c-\frac{7}{2}}\Omega_{i\nu}(1/\sqrt{\eta})\Gamma \left(\frac{1}{2} (- a- b+ c- i \nu +5/2)\right)\nonumber \\ 
&\times \Gamma \left(\frac{1}{2} ( a+ b+ c- i \nu +5/2)\right) \Gamma \left(\frac{1}{2} (- a- b+ c+ i \nu +5/2)\right) \Gamma \left(\frac{1}{2} ( a+ b+ c+ i \nu +5/2)\right)\beta(\nu).
\end{align}

This agrees exactly with the impact parameter calculation (\ref{eq:impactfinal}) after plugging in the definition for $V$ and setting $d=4$.
\subsection{General d}
Now we will claim that the impact parameter formalism works in odd dimensions for general blocks when we integrate the blocks against functions of the following form:
\bea
f(h,\bar{h},c,d,m)=2^{-2(h+\bar{h})}(h\bar{h})^{c}(h^{2}-\bar{h}^{2})^{d/2-1}(h^{2}+\bar{h}^{2})^m, \label{eq:OddAnsatz}
\eea

where $c$ is an arbitrary number, and $m$ is an integer. If this holds, this would prove the impact parameter formulas yield the correct result in general dimensions when the exchanged operator has an integer twist. The restriction to integer $m$ arises due to the use of the quadratic Casimir operator in the proof, which has eigenvalue $h^{2}+\bar{h}^{2}$.

Integrating (\ref{eq:OddAnsatz}) for $m=0$ against (\ref{eq:genImp}) we obtain:
\begin{align}
I_{impact}|_{m=0}&=\int dh d\bar{h} f(h,\bar{h},c,d,0)\mathcal{I}^{t,(a,b)}_{h,\bar{h}} \nonumber
\\=&\frac{1}{\pi }2^{-d-2} \Gamma (\frac{1}{2} (- a- b+ c+3/2)) \Gamma (\frac{1}{2} ( a+ b+ c+3/2))  \nonumber
\\ &\times \Gamma (\frac{1}{2} (- a- b+ c+ d-1/2)) \Gamma (\frac{1}{2} ( a+ b+ c+ d-1/2))(|p||q|)^{1/2 - a - b - c - d}.
\end{align}

For odd dimensions we do not have a simple closed form for the conformal blocks, but identity matching tells us that when doing the t-channel expansion for $\<\f_{1}\f_{1}\f_{2}\f_{2}\>$ we have:
\begin{align}
\int dh d\bar{h}& G^{(d),a,a}_{h,\bar{h}}(z,\bar{z})(h\bar{h})^{\frac{1}{2}-d+\Delta_{1}+\Delta_{2}}2^{-2(h+\bar{h})}(h^{2}-\bar{h}^{2})^{d/2-1}\nonumber\\ & =\frac{1}{(z\bar{z})^{\Delta_{2}}}\pi^{-1}2^{-2-d}\Gamma(\Delta_{1})\Gamma(1-d/2+\Delta_{1})\Gamma(\Delta_{2})\Gamma(1-d/2+\Delta_{2}),
\end{align}
where all we have done is move the constant pieces of the MFT OPE coefficients to the right hand side and used $a=\frac{1}{2}(\Delta_{2}-\Delta_{1})$. Now we can do the following trick: first change variables from $(\Delta_{1},\Delta_{2})$ to $(a,c)$, with $a$ given above and $c=\frac{1}{2}-d+\Delta_{1}+\Delta_{2}$. Then we use the fact that the t-channel conformal blocks in the s-channel Regge limit are a function of $a+b$ to make the replacement $a\rightarrow \frac{1}{2}(a+b)$ everywhere on the right hand side. This yields:
\begin{align}
\int dh d\bar{h}& f(h,\bar{h},c,d,0) G^{(d),a,b}_{h,\bar{h}}(z,\bar{z})\\
=&\frac{1}{\pi }2^{-d-2} \Gamma \left(\frac{1}{4} (-2 a-2 b+2 c+3)\right) \Gamma \left(\frac{1}{4} (2 a+2 b+2 c+3)\right) \nonumber \\ &\Gamma \left(\frac{1}{4} (-2 a-2 b+2 c+2 d-1)\right) \Gamma \left(\frac{1}{4} (2 a+2 b+2 c+2 d-1)\right) (z \bar{z})^{\frac{1}{4} (-2 a-2 b-2 c-2 d+1)}.
\end{align}
Finally we note that $|p||q|=\sqrt{z\bar{z}}$, and this proves the equality for the integrated blocks for $m=0$.

Now we can consider $m\neq0$. For $m\geq0$ and integer we can use the fact that both the conformal and impact parameter blocks are eigenfunctions of the quadratic Casimir with eigenvalues $h^{2}+\bar{h}^{2}$. By acting with the quadratic Casimir $D^{t}_{2}$ we can generate higher powers of $m$. Since the two integrals match for the base case, $m=0$, they hold for all positive, integer $m$ as well.

For $m<0$ the only subtlety is if the integral
\bea
F(z,\bar{z},c,d,m)=\int dh d\bar{h}(\mathcal{I}^{t}_{h,\bar{h}}(z,\bar{z})-G_{h,\bar{h}}(z,\bar{z}))f(h,\bar{h},c,d,m)
\eea
lies in the kernel of $D^{t}_{2}$ and is equally as divergent as the integrated blocks. As a reminder, this differential operator is given by:
\bea
D^{t}=z\partial^{2}+\bar{z}\bar{\partial}^{2}+(a+b+1)(\partial+\bar{\partial})+\frac{d-2}{z-\bar{z}}(z\partial-\bar{\partial}).
\eea
As long as $a+b\neq -1$ or  $2(a+b)\neq -1$, such zero eigenfunctions must be subleading and there will be no subtlety. So barring this complication, if $F(z,\bar{z},c,d,m)\neq 0$ for $m\leq0$ and integral, then repeatedly acting with the quadratic casimir would imply $F(z,\bar{z},c,d,0)\neq 0$, which is a contradiction. 

In summary we have shown the conformal and impact parameter blocks agree when integrated against (\ref{eq:OddAnsatz}) for $m$ an arbitrary integer. The anomalous dimensions found in \cite{Cornalba:2007zb} can be decomposed into such functions when the exchanged operator has integer twist, so the standard conformal block decomposition will also give the same answer. Proving their formulas for general dimensions and general twist remains an open question.

\section{Double-Trace Operators in an Example}
\label{App:DoubleTrace}

In this appendix we analyze a simple example in detail to illustrate
the method in section \ref{sec:GravityLimit}. We consider a 4d CFT and the four-point function
$\langle\psi\phi\phi\psi\rangle$ of scalars. We will set the external dimensions
to special values, in particular $\Delta_{\psi}=2$ and $\Delta_{\phi}=\frac{3}{2}$.
This makes all the calculations easy while still maintaining the general
features described in section \ref{sec:GravityLimit}.

Consider the $\psi\psi$ channel stress tensor block in the
Regge limit: 
\begin{equation}
A(\sigma,\eta)\supset C_{\phi\phi T}C_{\psi\psi T}g_{T}=i\pi\frac{960}{C_{T}}\frac{\eta}{\sigma}\frac{1}{1-\eta}\label{eq:SpecialStressTensorBlock}.
\end{equation}
There are also contributions from spin-2 $[\psi\psi]$ and $[\phi\phi]$
double-trace operators that appear at the same order in $\sigma^{-1}$.

In the $\psi\phi$ channel, $A(\sigma,\eta)$ is reproduced by the sum over large
spin, large twist double-trace operators. We can approximate the OPE
sum as an integral: 
\begin{equation}
A(\sigma,\eta)=-16\pi i\sigma^{2}\eta^{\frac{3}{2}}\frac{1}{1-\eta}\int_{0}^{\infty}dh\int_{0}^{h}d\bar{h}\left(e^{-2\sqrt{\sigma}\left(\sqrt{\eta}h+\bar{h}\right)}-e^{-2\sqrt{\sigma}\left(h+\sqrt{\eta}\bar{h}\right)}\right)\left(h^{2}-\bar{h}^{2}\right)\gamma_{h,\bar{h}},\label{eq:SpecialCrossedChannel}
\end{equation}
where we have already plugged in the mean field theory OPE coefficients
that are fixed by identity matching:
\begin{equation}
P^{MFT}\rightarrow2^{-2(h+\bar{h}-1)+5}(h^{2}-\bar{h}^{2}).
\end{equation}
We also used that the $\psi\phi$ channel conformal blocks simplify
to: 
\begin{equation}
g_{h,\bar{h}}^{d=4}=4^{h+\bar{h}-2}\frac{1}{\sigma(1-\eta)}\left(e^{-2\sqrt{\sigma}\left(\sqrt{\eta}h+\bar{h}\right)}-e^{-2\sqrt{\sigma}\left(h+\sqrt{\eta}\bar{h}\right)}\right).
\end{equation}
We first concentrate on matching the stress tensor contribution using the methods of section \ref{sec:GravityLimit} to
$A(\sigma,\eta)$ and obtain a simple solution to this equation:\footnote{The $h^{2}-\bar{h}^{2}$ factor cancels the one from (\ref{eq:SpecialCrossedChannel}).
The total power of $h$ and $\bar{h}$ is fixed to reproduce the $\sigma^{-1}$
behavior in $A(\sigma,\eta)$. The particular power $\bar{h}^{4}h^{0}$
is fixed by the power $\eta$ in (\ref{eq:SpecialStressTensorBlock}). }

\begin{equation}
\gamma_{h,\bar{h}}^{2,\frac{3}{2}}=-\frac{160}{C_{T}}\frac{\bar{h}^{4}}{h^{2}-\bar{h}^{2}}.
\end{equation}
We can plug this solution back into (\ref{eq:SpecialCrossedChannel})
to obtain the full amplitude: 
\begin{equation}
A(\sigma,\eta)=i\pi\frac{960}{C_{T}}\frac{\eta}{\sigma}\frac{1}{(1+\sqrt{\eta})^{6}}.\label{eq:SpecialA}
\end{equation}
Note that it is also regular
when we send $\eta\rightarrow1$, which corresponds to a vanishing impact
parameter.\footnote{Recall that we are sending $\Delta_{gap}\rightarrow\infty$ first.
In other words, we take $1\gg1-\eta\gg\frac{1}{\Delta_{gap}}\gg\frac{1}{C_{T}}$. }

We verify that the difference between (\ref{eq:SpecialStressTensorBlock})
and (\ref{eq:SpecialA}) is exactly reproduced by the appropriate
$\psi\psi\rightarrow\f\f$ channel double-trace operators that dress the stress
tensor block into a bulk Witten diagram. This implies that such double-trace contributions, being fixed by the single-trace data, do not affect the anomalous dimensions even when the contribute at the same order as the single-trace block in the Regge limit.
To do this, we first recall the form of $\alpha(\nu)$ in the gravity limit (\ref{eq:alpha4d}):
\begin{equation}
\alpha(\nu)=\frac{4\pi}{N^{2}}\frac{1}{4+\nu^{2}}\frac{\Gamma(\Delta_{1}+i\nu/2)\Gamma(\Delta_{1}-i\nu/2)\Gamma(\Delta_{2}+i\nu/2)\Gamma(\Delta_{2}-i\nu/2)}{\Gamma(\Delta_{1})\Gamma(\Delta_{1}-1)\Gamma(\Delta_{2})\Gamma(\Delta_{2}-1)} .
\end{equation}
This contains poles in $\nu$ that correpond to the stress tensor
at $i\nu=2$ as well as $[\psi\psi]$ and $[\phi\phi]$ spin-2 double
trace operators at $i\nu=2\Delta_{1,2}+2n$. For our particular choice
of external dimensions, this function simplifies to
\begin{equation}
\left.\alpha(\nu)\right|_{\Delta_{1}=2,\ \Delta_{2}=\frac{3}{2}}=i\pi^{4}\frac{120}{C_{T}}\frac{\nu\left(\nu^{2}+1\right)}{\sinh(\pi\nu)}.
\end{equation}
We then evaluate (\ref{eq:4ptRegge}), obtaining
\begin{equation}
A(\sigma,\eta)=i\pi\frac{960}{C_{T}}\frac{\eta}{\sigma}\frac{1}{\left(1+\sqrt{\eta}\right)^{6}}.
\end{equation}
This indeed agrees with (\ref{eq:SpecialA}). In evaluating this integral,
we used the harmonic functions on hyperbolic space (\ref{eq:4dReggeBlock}) as
well as the integral
\begin{equation}
\int_{-\infty}^{\infty}d\nu\nu^{2}(1+\nu^{2})\frac{\sin\rho\nu}{\sinh\nu}=48\frac{\sinh^{6}\frac{\rho}{2}}{\sinh^{5}\rho},
\end{equation}
in which $\rho=-\frac{1}{2}\log\eta$. 

Alternatively, we can evaluate the integral (\ref{eq:4ptRegge})
by summing over the residues. For example, the contributions corresponding
to the $[\psi\psi]$ operators are
\begin{equation}
2\pi i\mathcal{R}\left[\alpha(\nu)\right]_{i\nu=2\Delta_{1}+2n}=\frac{320}{\pi C_{T}}\frac{2\pi^{2}(-1)^{-n}\Gamma\left(n+2\Delta_{1}\right)\Gamma\left(-n-\Delta_{1}+\Delta_{2}\right)\Gamma\left(n+\Delta_{1}+\Delta_{2}\right)}{n!\left(\Delta_{1}+n-1\right)\left(\Delta_{1}+n+1\right)}.
\end{equation}
On the other hand, the stress tensor pole yields
\begin{equation}
2\pi i\mathcal{R}\left[\alpha(\nu)\right]_{i\nu=2}=\frac{3}{2}\pi^{2}\frac{320}{C_{T}}.
\end{equation}
We sum over these contributions to obtain $A(\sigma,\eta)$:
\begin{equation}
A(\sigma,\eta)=\frac{1}{\sigma\sqrt{\eta}}{\displaystyle \sum_{\Im \nu_{i}>0}}2\pi i\mathcal{R}\left[\alpha(\nu)\Omega_{i\nu}\left(\frac{-\log\eta}{2}\right)\right]_{\nu=\nu_{i}},
\end{equation}
where the poles are located at
\begin{equation}
\{-i\nu_{i}\}=\left\{ 2,\ 2\Delta_{1}+2n,\ 2\Delta_{2}+2n,\hspace{1em}n\ge0\right\} .
\end{equation}
In the special case we considered, this sum can be done by Mathematica
in closed form. The result is
\begin{equation}
A(\sigma,\eta)=i\pi\frac{960}{C_{T}}\frac{\eta}{\sigma}\frac{1}{\left(1+\sqrt{\eta}\right)^{6}},
\end{equation}
which again agrees with (\ref{eq:SpecialA}).

\section{Partial Wave Positivity}
\label{app:PartialWavePositivity}
In this appendix, we present an alternative argument for the negativity of anomalous dimensions that does not require an explicit specific solution of the bootstrap equations. However, this argument requires that a specific type of s-channel partial wave sum defines an analytic function in an upper half disk region of the complex $\sigma$ plane.
The idea is to focus on a partial contribution to the four-point function corresponding to a narrow wedge for $\bar{h}/h$ centered at the value $r$ of small width $\delta$:
\bea
G_{r,\delta}(z,\bar{z})=\underset{r-\delta\leq \bar{h}/h \leq r+\delta}{\sum}P_{\mathcal{O}}g_{\mathcal{O}}(z,\bar{z})=c(r,\delta)+\frac{i}{\s C_{T}}f_{r,\delta}(\eta) .\label{eq:sumrepRestriction}
\eea
Here we take $r$ to lie within the unit interval, $0<r<1$ and $\delta$ is an infinitesimal parameter.\footnote{We also restrict to $j=h-\bar{h}>2$ so we can later ignore contributions of homogenous solutions (i.e. contact diagrams in the bulk) to the anomalous dimensions of operators with spin $\leq 2$.} The function $c(r,\delta)$ is some positive number which corresponds to the contribution of the restricted s-channel sum to identity matching. 

Each s-channel conformal block has the right positivity properties (\ref{eq:PositiveCoeff}) necessary for the chaos bound. Using the general correspondence (\ref{eq:saddles}), the $\sigma^{-1}$ term from the spin-2 operators in the $\psi\psi$ OPE implies that the anomalous dimensions in the $\psi\phi$ channel grow like $h\bar{h}$. Therefore, we can write them in the following form: 
\bea
\gamma_{h,\bar{h}}= \frac{h\bar{h}}{C_{T}}(\beta(\bar{h}/h)+\mathcal{O}(h^{-1})),\hspace{1cm}C_T\gg\Delta_{gap}\gg h\sim\bar{h}\gg1.
\label{eq:AnomalousDimForm}
\eea
for some function $\beta(\bar{h}/h)$.

Finally, we will write down an integral expression for the $C_{T}^{-1}$ piece of (\ref{eq:sumrepRestriction}) in the Regge limit. Taking $\delta\rightarrow0$ so $\bar{h}/h\approx r$ and introducing the variable $s=4h\bar{h}$, we find:
\bea
\lim_{\delta\rightarrow 0}\frac{1}{2\delta}f_{r,\delta}(z,\bar{z})=-\pi\s \frac{\beta(r)}{r}\int_{0}^{\infty} ds \frac{s}{32r}P^{MFT}_{h,rh}g_{h,rh}(z,\bar{z}). \label{eq:int_f}
\eea
The factor of $h\bar{h}=s/4$ from the anomalous dimensions leads to a $\s^{-1}$ divergence which cancels the explicit $\s$ in (\ref{eq:int_f}). The integrand of (\ref{eq:int_f}) is explicitly positive: the conformal blocks are positive since we are in a reflection positive configuration and the OPE coefficients squared are positive by unitarity. The sign of the function $f_{r,\delta}(\eta)$ is thus determined by the sign of $\beta(r)$. 

The chaos bound, assuming it can be applied to each restricted sum, then implies that $\beta(r)\le0$ for $0<r<1$. Together with (\ref{eq:AnomalousDimForm}), we then get the negativity of anomalous dimensions $\gamma_{h,\bar{h}}\le0$ for $h\sim\bar{h}\gg1$. In order for this argument to work, it is important that the projected sums~(\ref{eq:sumrepRestriction}) define an analytic function in an appropriate region of the complex $\sigma$ plane which is bounded by $c(r,\delta)$ along the real $\sigma$ line in both the $s$-channel regime $\sigma > 0$ and the $u$-channel regime $\sigma < 0$. While this can be explicitly checked in cases where the double-trace operators dominate, we have not yet understood a rigorous argument establishing this property in the $u$-channel regime when $|\sigma| \lesssim 1/C_T$. We hope this gap can be overcome in future work.

\bibliography{Biblio}{}

\providecommand{\href}[2]{#2}\begingroup\raggedright\begin{thebibliography}{10}

\bibitem{Ferrara:1973yt}
S.~Ferrara, A.~F. Grillo, and R.~Gatto, ``{Tensor representations of conformal
  algebra and conformally covariant operator product expansion},''
\href{http://dx.doi.org/10.1016/0003-4916(73)90446-6}{{\em Annals Phys.}
  {\bfseries 76} (1973) 161--188}.
%%CITATION = APNYA,76,161;%%.

\bibitem{Polyakov:1974gs}
A.~M. Polyakov, ``{Nonhamiltonian approach to conformal quantum field
  theory},''
{\em Zh. Eksp. Teor. Fiz.} {\bfseries 66} (1974) 23--42.
%%CITATION = ZETFA,66,23;%%.

\bibitem{Rattazzi:2008pe}
R.~Rattazzi, V.~S. Rychkov, E.~Tonni, and A.~Vichi, ``{Bounding scalar operator
  dimensions in 4D CFT},''
  \href{http://dx.doi.org/10.1088/1126-6708/2008/12/031}{{\em JHEP} {\bfseries
  12} (2008) 031},
\href{http://arxiv.org/abs/0807.0004}{{\ttfamily arXiv:0807.0004 [hep-th]}}.
%%CITATION = 0807.0004;%%.

\bibitem{ElShowk:2012ht}
S.~El-Showk, M.~F. Paulos, D.~Poland, S.~Rychkov, D.~Simmons-Duffin, and
  A.~Vichi, ``{Solving the 3D Ising Model with the Conformal Bootstrap},''
  \href{http://dx.doi.org/10.1103/PhysRevD.86.025022}{{\em Phys.Rev.}
  {\bfseries D86} (2012) 025022},
\href{http://arxiv.org/abs/1203.6064}{{\ttfamily arXiv:1203.6064 [hep-th]}}.
%%CITATION = ARXIV:1203.6064;%%.

\bibitem{El-Showk:2014dwa}
S.~El-Showk, M.~F. Paulos, D.~Poland, S.~Rychkov, D.~Simmons-Duffin, {\em et
  al.}, ``{Solving the 3d Ising Model with the Conformal Bootstrap II.
  c-Minimization and Precise Critical Exponents},''
  \href{http://dx.doi.org/10.1007/s10955-014-1042-7}{{\em J.Stat.Phys.}
  {\bfseries 157} (2014) 869},
\href{http://arxiv.org/abs/1403.4545}{{\ttfamily arXiv:1403.4545 [hep-th]}}.
%%CITATION = ARXIV:1403.4545;%%.

\bibitem{Kos:2014bka}
F.~Kos, D.~Poland, and D.~Simmons-Duffin, ``{Bootstrapping Mixed Correlators in
  the 3D Ising Model},'' \href{http://dx.doi.org/10.1007/JHEP11(2014)109}{{\em
  JHEP} {\bfseries 1411} (2014) 109},
\href{http://arxiv.org/abs/1406.4858}{{\ttfamily arXiv:1406.4858 [hep-th]}}.
%%CITATION = ARXIV:1406.4858;%%.

\bibitem{Simmons-Duffin:2015qma}
D.~Simmons-Duffin, ``{A Semidefinite Program Solver for the Conformal
  Bootstrap},'' \href{http://dx.doi.org/10.1007/JHEP06(2015)174}{{\em JHEP}
  {\bfseries 06} (2015) 174},
\href{http://arxiv.org/abs/1502.02033}{{\ttfamily arXiv:1502.02033 [hep-th]}}.
%%CITATION = ARXIV:1502.02033;%%.

\bibitem{Kos:2016ysd}
F.~Kos, D.~Poland, D.~Simmons-Duffin, and A.~Vichi, ``{Precision islands in the
  Ising and O(N) models},''
  \href{http://dx.doi.org/10.1007/JHEP08(2016)036}{{\em JHEP} {\bfseries 08}
  (2016) 036},
\href{http://arxiv.org/abs/1603.04436}{{\ttfamily arXiv:1603.04436 [hep-th]}}.
%%CITATION = ARXIV:1603.04436;%%.

\bibitem{Fitzpatrick:2012yx}
A.~L. Fitzpatrick, J.~Kaplan, D.~Poland, and D.~Simmons-Duffin, ``{The Analytic
  Bootstrap and AdS Superhorizon Locality},''
  \href{http://dx.doi.org/10.1007/JHEP12(2013)004}{{\em JHEP} {\bfseries 1312}
  (2013) 004},
\href{http://arxiv.org/abs/1212.3616}{{\ttfamily arXiv:1212.3616 [hep-th]}}.
%%CITATION = ARXIV:1212.3616;%%.

\bibitem{Komargodski:2012ek}
Z.~Komargodski and A.~Zhiboedov, ``{Convexity and Liberation at Large Spin},''
  \href{http://dx.doi.org/10.1007/JHEP11(2013)140}{{\em JHEP} {\bfseries 1311}
  (2013) 140},
\href{http://arxiv.org/abs/1212.4103}{{\ttfamily arXiv:1212.4103 [hep-th]}}.
%%CITATION = ARXIV:1212.4103;%%.

\bibitem{Li:2015rfa}
D.~Li, D.~Meltzer, and D.~Poland, ``{Non-Abelian Binding Energies from the
  Lightcone Bootstrap},'' \href{http://dx.doi.org/10.1007/JHEP02(2016)149}{{\em
  JHEP} {\bfseries 02} (2016) 149},
\href{http://arxiv.org/abs/1510.07044}{{\ttfamily arXiv:1510.07044 [hep-th]}}.
%%CITATION = ARXIV:1510.07044;%%.

\bibitem{Li:2015itl}
D.~Li, D.~Meltzer, and D.~Poland, ``{Conformal Collider Physics from the
  Lightcone Bootstrap},'' \href{http://dx.doi.org/10.1007/JHEP02(2016)143}{{\em
  JHEP} {\bfseries 02} (2016) 143},
\href{http://arxiv.org/abs/1511.08025}{{\ttfamily arXiv:1511.08025 [hep-th]}}.
%%CITATION = ARXIV:1511.08025;%%.

\bibitem{Hofman:2016awc}
D.~M. Hofman, D.~Li, D.~Meltzer, D.~Poland, and F.~Rejon-Barrera, ``{A Proof of
  the Conformal Collider Bounds},''
  \href{http://dx.doi.org/10.1007/JHEP06(2016)111}{{\em JHEP} {\bfseries 06}
  (2016) 111},
\href{http://arxiv.org/abs/1603.03771}{{\ttfamily arXiv:1603.03771 [hep-th]}}.
%%CITATION = ARXIV:1603.03771;%%.

\bibitem{Fitzpatrick:2016ive}
A.~L. Fitzpatrick, J.~Kaplan, D.~Li, and J.~Wang, ``{On information loss in
  AdS$_{3}$/CFT$_{2}$},'' \href{http://dx.doi.org/10.1007/JHEP05(2016)109}{{\em
  JHEP} {\bfseries 05} (2016) 109},
\href{http://arxiv.org/abs/1603.08925}{{\ttfamily arXiv:1603.08925 [hep-th]}}.
%%CITATION = ARXIV:1603.08925;%%.

\bibitem{Dyer:2016pou}
E.~Dyer and G.~Gur-Ari, ``{2D CFT Partition Functions at Late Times},''
\href{http://arxiv.org/abs/1611.04592}{{\ttfamily arXiv:1611.04592 [hep-th]}}.
%%CITATION = ARXIV:1611.04592;%%.

\bibitem{Chen:2017yze}
H.~Chen, C.~Hussong, J.~Kaplan, and D.~Li, ``{A Numerical Approach to Virasoro
  Blocks and the Information Paradox},''
\href{http://arxiv.org/abs/1703.09727}{{\ttfamily arXiv:1703.09727 [hep-th]}}.
%%CITATION = ARXIV:1703.09727;%%.

\bibitem{Heemskerk:2009pn}
I.~Heemskerk, J.~Penedones, J.~Polchinski, and J.~Sully, ``{Holography from
  Conformal Field Theory},''
  \href{http://dx.doi.org/10.1088/1126-6708/2009/10/079}{{\em JHEP} {\bfseries
  0910} (2009) 079},
\href{http://arxiv.org/abs/0907.0151}{{\ttfamily arXiv:0907.0151 [hep-th]}}.
%%CITATION = ARXIV:0907.0151;%%.

\bibitem{Heemskerk:2010ty}
I.~Heemskerk and J.~Sully, ``{More Holography from Conformal Field Theory},''
  \href{http://dx.doi.org/10.1007/JHEP09(2010)099}{{\em JHEP} {\bfseries 1009}
  (2010) 099},
\href{http://arxiv.org/abs/1006.0976}{{\ttfamily arXiv:1006.0976 [hep-th]}}.
%%CITATION = ARXIV:1006.0976;%%.

\bibitem{Afkhami-Jeddi:2016ntf}
N.~Afkhami-Jeddi, T.~Hartman, S.~Kundu, and A.~Tajdini, ``{Einstein gravity
  3-point functions from conformal field theory},''
\href{http://arxiv.org/abs/1610.09378}{{\ttfamily arXiv:1610.09378 [hep-th]}}.
%%CITATION = ARXIV:1610.09378;%%.

\bibitem{Caron-Huot:2017vep}
S.~Caron-Huot, ``{Analyticity in Spin in Conformal Theories},''
\href{http://arxiv.org/abs/1703.00278}{{\ttfamily arXiv:1703.00278 [hep-th]}}.
%%CITATION = ARXIV:1703.00278;%%.

\bibitem{Alday:2014tsa}
L.~F. Alday, A.~Bissi, and T.~Lukowski, ``{Lessons from crossing symmetry at
  large N},'' \href{http://dx.doi.org/10.1007/JHEP06(2015)074}{{\em JHEP}
  {\bfseries 06} (2015) 074},
\href{http://arxiv.org/abs/1410.4717}{{\ttfamily arXiv:1410.4717 [hep-th]}}.
%%CITATION = ARXIV:1410.4717;%%.

\bibitem{Alday:2016htq}
L.~F. Alday and A.~Bissi, ``{Unitarity and positivity constraints for CFT at
  large central charge},''
\href{http://arxiv.org/abs/1606.09593}{{\ttfamily arXiv:1606.09593 [hep-th]}}.
%%CITATION = ARXIV:1606.09593;%%.

\bibitem{Aharony:2016dwx}
O.~Aharony, L.~F. Alday, A.~Bissi, and E.~Perlmutter, ``{Loops in AdS from
  Conformal Field Theory},''
\href{http://arxiv.org/abs/1612.03891}{{\ttfamily arXiv:1612.03891 [hep-th]}}.
%%CITATION = ARXIV:1612.03891;%%.

\bibitem{Rastelli:2016nze}
L.~Rastelli and X.~Zhou, ``{Mellin amplitudes for $AdS_5\times S^5$},''
  \href{http://dx.doi.org/10.1103/PhysRevLett.118.091602}{{\em Phys. Rev.
  Lett.} {\bfseries 118} no.~9, (2017) 091602},
\href{http://arxiv.org/abs/1608.06624}{{\ttfamily arXiv:1608.06624 [hep-th]}}.
%%CITATION = ARXIV:1608.06624;%%.

\bibitem{Perlmutter:2016pkf}
E.~Perlmutter, ``{Bounding the Space of Holographic CFTs with Chaos},''
  \href{http://dx.doi.org/10.1007/JHEP10(2016)069}{{\em JHEP} {\bfseries 10}
  (2016) 069},
\href{http://arxiv.org/abs/1602.08272}{{\ttfamily arXiv:1602.08272 [hep-th]}}.
%%CITATION = ARXIV:1602.08272;%%.

\bibitem{Brower:2006ea}
R.~C. Brower, J.~Polchinski, M.~J. Strassler, and C.-I. Tan, ``{The Pomeron and
  gauge/string duality},''
  \href{http://dx.doi.org/10.1088/1126-6708/2007/12/005}{{\em JHEP} {\bfseries
  12} (2007) 005},
\href{http://arxiv.org/abs/hep-th/0603115}{{\ttfamily arXiv:hep-th/0603115
  [hep-th]}}.
%%CITATION = HEP-TH/0603115;%%.

\bibitem{Cornalba:2007fs}
L.~Cornalba, ``{Eikonal methods in AdS/CFT: Regge theory and multi-reggeon
  exchange},''
\href{http://arxiv.org/abs/0710.5480}{{\ttfamily arXiv:0710.5480 [hep-th]}}.
%%CITATION = ARXIV:0710.5480;%%.

\bibitem{Cornalba:2007zb}
L.~Cornalba, M.~S. Costa, and J.~Penedones, ``{Eikonal approximation in
  AdS/CFT: Resumming the gravitational loop expansion},''
  \href{http://dx.doi.org/10.1088/1126-6708/2007/09/037}{{\em JHEP} {\bfseries
  09} (2007) 037},
\href{http://arxiv.org/abs/0707.0120}{{\ttfamily arXiv:0707.0120 [hep-th]}}.
%%CITATION = ARXIV:0707.0120;%%.

\bibitem{Cornalba:2006xk}
L.~Cornalba, M.~S. Costa, J.~Penedones, and R.~Schiappa, ``{Eikonal
  Approximation in AdS/CFT: From Shock Waves to Four-Point Functions},''
  \href{http://dx.doi.org/10.1088/1126-6708/2007/08/019}{{\em JHEP} {\bfseries
  08} (2007) 019},
\href{http://arxiv.org/abs/hep-th/0611122}{{\ttfamily arXiv:hep-th/0611122
  [hep-th]}}.
%%CITATION = HEP-TH/0611122;%%.

\bibitem{Cornalba:2006xm}
L.~Cornalba, M.~S. Costa, J.~Penedones, and R.~Schiappa, ``{Eikonal
  Approximation in AdS/CFT: Conformal Partial Waves and Finite N Four-Point
  Functions},'' \href{http://dx.doi.org/10.1016/j.nuclphysb.2007.01.007}{{\em
  Nucl. Phys.} {\bfseries B767} (2007) 327--351},
\href{http://arxiv.org/abs/hep-th/0611123}{{\ttfamily arXiv:hep-th/0611123
  [hep-th]}}.
%%CITATION = HEP-TH/0611123;%%.

\bibitem{Costa:2012cb}
M.~S. Costa, V.~Goncalves, and J.~Penedones, ``{Conformal Regge theory},''
  \href{http://dx.doi.org/10.1007/JHEP12(2012)091}{{\em JHEP} {\bfseries 1212}
  (2012) 091},
\href{http://arxiv.org/abs/1209.4355}{{\ttfamily arXiv:1209.4355 [hep-th]}}.
%%CITATION = ARXIV:1209.4355;%%.

\bibitem{Maldacena:2015waa}
J.~Maldacena, S.~H. Shenker, and D.~Stanford, ``{A bound on chaos},''
  \href{http://dx.doi.org/10.1007/JHEP08(2016)106}{{\em JHEP} {\bfseries 08}
  (2016) 106},
\href{http://arxiv.org/abs/1503.01409}{{\ttfamily arXiv:1503.01409 [hep-th]}}.
%%CITATION = ARXIV:1503.01409;%%.

\bibitem{Camanho:2014apa}
X.~O. Camanho, J.~D. Edelstein, J.~Maldacena, and A.~Zhiboedov, ``{Causality
  Constraints on Corrections to the Graviton Three-Point Coupling},''
  \href{http://dx.doi.org/10.1007/JHEP02(2016)020}{{\em JHEP} {\bfseries 02}
  (2016) 020},
\href{http://arxiv.org/abs/1407.5597}{{\ttfamily arXiv:1407.5597 [hep-th]}}.
%%CITATION = ARXIV:1407.5597;%%.

\bibitem{Caron-Huot:2016icg}
S.~Caron-Huot, Z.~Komargodski, A.~Sever, and A.~Zhiboedov, ``{Strings from
  Massive Higher Spins: The Asymptotic Uniqueness of the Veneziano
  Amplitude},''
\href{http://arxiv.org/abs/1607.04253}{{\ttfamily arXiv:1607.04253 [hep-th]}}.
%%CITATION = ARXIV:1607.04253;%%.

\bibitem{Cardona:2016ymb}
C.~Cardona, Y.-t. Huang, and T.-H. Tsai, ``{On the linearity of Regge
  trajectory at large transfer energy},''
\href{http://arxiv.org/abs/1611.05797}{{\ttfamily arXiv:1611.05797 [hep-th]}}.
%%CITATION = ARXIV:1611.05797;%%.

\bibitem{Hofman:2008ar}
D.~M. Hofman and J.~Maldacena, ``{Conformal collider physics: Energy and charge
  correlations},'' \href{http://dx.doi.org/10.1088/1126-6708/2008/05/012}{{\em
  JHEP} {\bfseries 05} (2008) 012},
\href{http://arxiv.org/abs/0803.1467}{{\ttfamily arXiv:0803.1467 [hep-th]}}.
%%CITATION = ARXIV:0803.1467;%%.

\bibitem{Hartman:2015lfa}
T.~Hartman, S.~Jain, and S.~Kundu, ``{Causality Constraints in Conformal Field
  Theory},'' \href{http://dx.doi.org/10.1007/JHEP05(2016)099}{{\em JHEP}
  {\bfseries 05} (2016) 099},
\href{http://arxiv.org/abs/1509.00014}{{\ttfamily arXiv:1509.00014 [hep-th]}}.
%%CITATION = ARXIV:1509.00014;%%.

\bibitem{Hartman:2016dxc}
T.~Hartman, S.~Jain, and S.~Kundu, ``{A New Spin on Causality Constraints},''
  \href{http://dx.doi.org/10.1007/JHEP10(2016)141}{{\em JHEP} {\bfseries 10}
  (2016) 141},
\href{http://arxiv.org/abs/1601.07904}{{\ttfamily arXiv:1601.07904 [hep-th]}}.
%%CITATION = ARXIV:1601.07904;%%.

\bibitem{Komargodski:2016gci}
Z.~Komargodski, M.~Kulaxizi, A.~Parnachev, and A.~Zhiboedov, ``{Conformal Field
  Theories and Deep Inelastic Scattering},''
  \href{http://dx.doi.org/10.1103/PhysRevD.95.065011}{{\em Phys. Rev.}
  {\bfseries D95} no.~6, (2017) 065011},
\href{http://arxiv.org/abs/1601.05453}{{\ttfamily arXiv:1601.05453 [hep-th]}}.
%%CITATION = ARXIV:1601.05453;%%.

\bibitem{Hartman:2016lgu}
T.~Hartman, S.~Kundu, and A.~Tajdini, ``{Averaged Null Energy Condition from
  Causality},''
\href{http://arxiv.org/abs/1610.05308}{{\ttfamily arXiv:1610.05308 [hep-th]}}.
%%CITATION = ARXIV:1610.05308;%%.

\bibitem{Faulkner:2016mzt}
T.~Faulkner, R.~G. Leigh, O.~Parrikar, and H.~Wang, ``{Modular Hamiltonians for
  Deformed Half-Spaces and the Averaged Null Energy Condition},''
  \href{http://dx.doi.org/10.1007/JHEP09(2016)038}{{\em JHEP} {\bfseries 09}
  (2016) 038},
\href{http://arxiv.org/abs/1605.08072}{{\ttfamily arXiv:1605.08072 [hep-th]}}.
%%CITATION = ARXIV:1605.08072;%%.

\bibitem{Alday:2017gde}
L.~F. Alday, A.~Bissi, and E.~Perlmutter, ``{Holographic Reconstruction of AdS
  Exchanges from Crossing Symmetry},''
\href{http://arxiv.org/abs/1705.02318}{{\ttfamily arXiv:1705.02318 [hep-th]}}.
%%CITATION = ARXIV:1705.02318;%%.

\bibitem{KPZ2017}
M.~Kulaxizi, A.~Parnachev, and A.~Zhiboedov, ``{Bulk Phase Shift, CFT Regge
  Limit and Einstein Gravity},''
\href{http://arxiv.org/abs/1705.02934}{{\ttfamily arXiv:1705.02934 [hep-th]}}.
%%CITATION = ARXIV:1705.02934;%%.

\bibitem{Kaviraj:2015xsa}
A.~Kaviraj, K.~Sen, and A.~Sinha, ``{Universal anomalous dimensions at large
  spin and large twist},''
  \href{http://dx.doi.org/10.1007/JHEP07(2015)026}{{\em JHEP} {\bfseries 07}
  (2015) 026},
\href{http://arxiv.org/abs/1504.00772}{{\ttfamily arXiv:1504.00772 [hep-th]}}.
%%CITATION = ARXIV:1504.00772;%%.

\bibitem{Kaviraj:2015cxa}
A.~Kaviraj, K.~Sen, and A.~Sinha, ``{Analytic bootstrap at large spin},''
  \href{http://dx.doi.org/10.1007/JHEP11(2015)083}{{\em JHEP} {\bfseries 11}
  (2015) 083},
\href{http://arxiv.org/abs/1502.01437}{{\ttfamily arXiv:1502.01437 [hep-th]}}.
%%CITATION = ARXIV:1502.01437;%%.

\bibitem{Alday:2016njk}
L.~F. Alday, ``{Large Spin Perturbation Theory},''
\href{http://arxiv.org/abs/1611.01500}{{\ttfamily arXiv:1611.01500 [hep-th]}}.
%%CITATION = ARXIV:1611.01500;%%.

\bibitem{Cornalba:2008qf}
L.~Cornalba, M.~S. Costa, and J.~Penedones, ``{Eikonal Methods in AdS/CFT: BFKL
  Pomeron at Weak Coupling},''
  \href{http://dx.doi.org/10.1088/1126-6708/2008/06/048}{{\em JHEP} {\bfseries
  06} (2008) 048},
\href{http://arxiv.org/abs/0801.3002}{{\ttfamily arXiv:0801.3002 [hep-th]}}.
%%CITATION = ARXIV:0801.3002;%%.

\bibitem{Fitzpatrick:2010zm}
A.~Fitzpatrick, E.~Katz, D.~Poland, and D.~Simmons-Duffin, ``{Effective
  Conformal Theory and the Flat-Space Limit of AdS},''
  \href{http://dx.doi.org/10.1007/JHEP07(2011)023}{{\em JHEP} {\bfseries 1107}
  (2011) 023},
\href{http://arxiv.org/abs/1007.2412}{{\ttfamily arXiv:1007.2412 [hep-th]}}.
%%CITATION = ARXIV:1007.2412;%%.

\bibitem{Fitzpatrick:2011dm}
A.~L. Fitzpatrick and J.~Kaplan, ``{Unitarity and the Holographic S-Matrix},''
  \href{http://dx.doi.org/10.1007/JHEP10(2012)032}{{\em JHEP} {\bfseries 1210}
  (2012) 032},
\href{http://arxiv.org/abs/1112.4845}{{\ttfamily arXiv:1112.4845 [hep-th]}}.
%%CITATION = ARXIV:1112.4845;%%.

\bibitem{Rejon-Barrera:2015bpa}
F.~Rejon-Barrera and D.~Robbins, ``{Scalar-Vector Bootstrap},''
  \href{http://dx.doi.org/10.1007/JHEP01(2016)139}{{\em JHEP} {\bfseries 01}
  (2016) 139},
\href{http://arxiv.org/abs/1508.02676}{{\ttfamily arXiv:1508.02676 [hep-th]}}.
%%CITATION = ARXIV:1508.02676;%%.

\bibitem{Costa:2011dw}
M.~S. Costa, J.~Penedones, D.~Poland, and S.~Rychkov, ``{Spinning Conformal
  Blocks},'' \href{http://dx.doi.org/10.1007/JHEP11(2011)154}{{\em JHEP}
  {\bfseries 1111} (2011) 154},
\href{http://arxiv.org/abs/1109.6321}{{\ttfamily arXiv:1109.6321 [hep-th]}}.
%%CITATION = ARXIV:1109.6321;%%.

\bibitem{Brower:2007xg}
R.~C. Brower, M.~J. Strassler, and C.-I. Tan, ``{On The Pomeron at Large 't
  Hooft Coupling},''
  \href{http://dx.doi.org/10.1088/1126-6708/2009/03/092}{{\em JHEP} {\bfseries
  03} (2009) 092},
\href{http://arxiv.org/abs/0710.4378}{{\ttfamily arXiv:0710.4378 [hep-th]}}.
%%CITATION = ARXIV:0710.4378;%%.

\bibitem{Shenker:2014cwa}
S.~H. Shenker and D.~Stanford, ``{Stringy effects in scrambling},''
  \href{http://dx.doi.org/10.1007/JHEP05(2015)132}{{\em JHEP} {\bfseries 05}
  (2015) 132},
\href{http://arxiv.org/abs/1412.6087}{{\ttfamily arXiv:1412.6087 [hep-th]}}.
%%CITATION = ARXIV:1412.6087;%%.

\bibitem{Brower:2007qh}
R.~C. Brower, M.~J. Strassler, and C.-I. Tan, ``{On the eikonal approximation
  in AdS space},'' \href{http://dx.doi.org/10.1088/1126-6708/2009/03/050}{{\em
  JHEP} {\bfseries 03} (2009) 050},
\href{http://arxiv.org/abs/0707.2408}{{\ttfamily arXiv:0707.2408 [hep-th]}}.
%%CITATION = ARXIV:0707.2408;%%.

\bibitem{Amati:1987wq}
D.~Amati, M.~Ciafaloni, and G.~Veneziano, ``{Superstring Collisions at
  Planckian Energies},''
\href{http://dx.doi.org/10.1016/0370-2693(87)90346-7}{{\em Phys. Lett.}
  {\bfseries B197} (1987) 81}.
%%CITATION = PHLTA,B197,81;%%.

\bibitem{Amati:1987uf}
D.~Amati, M.~Ciafaloni, and G.~Veneziano, ``{Classical and Quantum Gravity
  Effects from Planckian Energy Superstring Collisions},''
\href{http://dx.doi.org/10.1142/S0217751X88000710}{{\em Int. J. Mod. Phys.}
  {\bfseries A3} (1988) 1615--1661}.
%%CITATION = IMPAE,A3,1615;%%.

\bibitem{Gubser:2002tv}
S.~S. Gubser, I.~R. Klebanov, and A.~M. Polyakov, ``{A Semiclassical limit of
  the gauge / string correspondence},''
  \href{http://dx.doi.org/10.1016/S0550-3213(02)00373-5}{{\em Nucl. Phys.}
  {\bfseries B636} (2002) 99--114},
\href{http://arxiv.org/abs/hep-th/0204051}{{\ttfamily arXiv:hep-th/0204051
  [hep-th]}}.
%%CITATION = HEP-TH/0204051;%%.

\bibitem{Alday:2015ewa}
L.~F. Alday and A.~Zhiboedov, ``{An Algebraic Approach to the Analytic
  Bootstrap},'' \href{http://dx.doi.org/10.1007/JHEP04(2017)157}{{\em JHEP}
  {\bfseries 04} (2017) 157},
\href{http://arxiv.org/abs/1510.08091}{{\ttfamily arXiv:1510.08091 [hep-th]}}.
%%CITATION = ARXIV:1510.08091;%%.

\bibitem{Alday:2015ota}
L.~F. Alday and A.~Zhiboedov, ``{Conformal Bootstrap With Slightly Broken
  Higher Spin Symmetry},''
  \href{http://dx.doi.org/10.1007/JHEP06(2016)091}{{\em JHEP} {\bfseries 06}
  (2016) 091},
\href{http://arxiv.org/abs/1506.04659}{{\ttfamily arXiv:1506.04659 [hep-th]}}.
%%CITATION = ARXIV:1506.04659;%%.

\bibitem{Simmons-Duffin:2016wlq}
D.~Simmons-Duffin, ``{The Lightcone Bootstrap and the Spectrum of the 3d Ising
  CFT},'' \href{http://dx.doi.org/10.1007/JHEP03(2017)086}{{\em JHEP}
  {\bfseries 03} (2017) 086},
\href{http://arxiv.org/abs/1612.08471}{{\ttfamily arXiv:1612.08471 [hep-th]}}.
%%CITATION = ARXIV:1612.08471;%%.

\end{thebibliography}\endgroup
\bibliographystyle{utphys}

\end{document}